\documentclass[twocolumn, tighten,]{aastex7}

\usepackage{amsmath}

\usepackage{graphicx}	
\usepackage{booktabs}
\usepackage{float}
\usepackage{caption}
\usepackage{multirow}
\usepackage{subcaption}              
\usepackage{tabularx}
\usepackage{graphicx}
\usepackage{adjustbox}
\usepackage{verbatim}
\usepackage{xcolor}

\newcommand{\her}{\textit{Herschel} }
\newcommand{\unet}{\textsc{U-Net} }
\newcommand{\disperse}{\textsc{DisPerSE}\ }
\newcommand{\filfinder}{\textsc{FilFinder}\ }
\newcommand{\getsf}{\textsc{getsf}\ }
\newcommand{\sutra}{\textsc{S\={u}tra}\,}
\newcommand{\nh}{N_{H_2}}
\newcommand{\ufil}{\textsc{dis:gsf}\,}

\newcommand{\tcm}[1]{\textcolor{black}{#1}} 

\accepted{May 21, 2026}
\submitjournal{Astronomical Journal (AJ), DOI: 10.3847/1538-3881/ae75db}

\begin{document}

\title{\sutra: An integrated framework for identification and characterization of filaments in the interstellar medium}

\author[orcid=0000-0002-9298-833X]{Shivam Kumaran} 
\affiliation{Space Applications Centre, ISRO, Ahmedabad, Gujarat, India, 380015}
\email[show]{kumaranshivam57@gmail.com}
\correspondingauthor{Shivam Kumaran}

\author[orcid=0009-0007-4978-6445]{Ushasi Bhowmick}
\affiliation{Space Applications Centre, ISRO, Ahmedabad, Gujarat, India, 380015}
\email{ushasibhowmick@gmail.com}

\author[orcid=0000-0002-1899-1020]{Vipin Kumar}
\affiliation{Space Applications Centre, ISRO, Ahmedabad, Gujarat, India, 380015}
\email{vipin0671@gmail.com}

\author[orcid=0009-0003-0445-0324]{Manish Chauhan}
\affiliation{Space Applications Centre, ISRO, Ahmedabad, Gujarat, India, 380015}
\email{manish.chauhan@outlook.in}

\author{Munn Vinayak Shukla}
\affiliation{Space Applications Centre, ISRO, Ahmedabad, Gujarat, India, 380015}
\email{munnvinayak@sac.isro.gov.in}

\author{Mehul R Pandya}
\affiliation{Space Applications Centre, ISRO, Ahmedabad, Gujarat, India, 380015}
\email{mrpandya@sac.isro.gov.in}

\begin{abstract}

Observations of the interstellar medium (ISM) at Far-infrared(FIR) and sub-millimetre (sub-mm) wavelengths reveal a complex filamentary structure of dust and gas, which plays a pivotal role in both low and high mass star formation. Large scale identification and characterization of filaments is crucial to establish a link between the ISM and the star formation. We present \sutra, a machine learning based framework that unifies filament identification and beam-scale physical characterization within a single automated pipeline. The framework employs a \unet\ architecture to perform supervised segmentation on column density maps and is trained on five nearby ($<$500\,pc) molecular clouds from the \her\ Gould Belt Survey (HGBS), using consensus skeletons constructed from the union of filaments identified by \disperse\ and \getsf. Rather than reproducing broad intensity-based masks, \sutra\ predicts crest-likelihood maps focused on filament spines.

Beyond identification, \sutra characterizes the filaments at the beam resolution by extracting radial profiles perpendicular to the crest and deriving local structural parameters. The framework provides a parameter-free, computationally efficient approach for consistent filaments identification and systematic investigation of their local properties \tcm{and shows stable behaviour across varying background conditions in controlled synthetic tests}. We demonstrate its application on selected regions from Aquila, Orion and Polaris molecular clouds, and compare the derived filament characteristics with those obtained using existing algorithms. \sutra robustly recovers filamentary structures consistent with cylindrical profiles, including in relatively low-intensity and low-contrast environments, making it well suited for both region-specific studies and large-scale statistical analyses of early-stage star formation and ISM structure.

\end{abstract}


\section{Introduction}
        
    Decades of observational and theoretical studies have shown that molecular clouds are the cradle of star formation \citep{McKee_2007}. While the composition and structure of molecular clouds have long been studied, sensitive observations highlight the presence of thread-like or filamentary structures \citep{Barnard_1927,Lynds_1962,Schneider_1979,Hartmann_2002,Stepnik_2003,Goldsmith_2008,Myers_2009}. With the advent of sensitive high resolution observations from the \textit{Herschel Space Observatory} \citep[\textit{Herschel,}][]{Pilbratt_2010}, it is observed that these structures are ubiquitous in molecular clouds \citep{Andre_2010,Molinari_2010,Motte_2010}. \textit{Herschel} observations also show most prestellar cores reside in dense filaments \citep{Andre_2010,Konyves_2015,Marsh_2016}, highlighting the importance of studying the distribution or structure of molecular gas in the interstellar medium (ISM) \citep{Andre_2014,Smith_2014}.
    
    Filaments exist across scales, from sub-parsec to kiloparsec in external galaxies \citep[see][]{Hacar_2023}. Studying the filament fragmentation holds the key to understanding early stages of star formation. Statistical studies on large samples of filaments suggest a characteristic width of $\sim$ 0.1$pc$ \citep{Arzoumanian_2011,Palmeirim_2013} and a Plummer-like radial profile \citep{Palmeirim_2013}. However, the origin and existence of a characteristic width is still debated \citep{Panopoulou_2017,Panopoulou_2022,Hacar_2023,andreTypicalWidth,mannfors}. Filaments are not only geometric features but are now recognized as the fundamental organizing structures of molecular clouds. Herschel surveys revealed that nearly all star-forming cores lie within parsec-scale filaments \citep{Andre_2014,Arzoumanian_2011}. Theoretical works suggest that turbulence and magnetic fields naturally generate elongated, self-gravitating structures\citep{Hacar_2023, Federrath}. Once formed, dense filaments can reach the critical line mass for gravitational instability, fragmenting into prestellar cores and thus setting the stage for clustered star formation. This underscores why robust filament identification and characterization is essential for understanding the star formation process. 
    
    To gain a deeper understanding of filament morphology and characteristics, several methodologies have been developed to identify filaments in column density (CD) maps, including but not limited to \filfinder \citep{filfinder_a,filfinder_b}, \getsf \citep{getsf}, \disperse \citep{Sousbie_2011}, template matching \citep{juvela} and Hessian matrix based methods\citep{2020MNRAS.492.5420S}. These broadly fall into two categories: local approaches and global approaches. Local methods, such as Hessian-based techniques, rely on thresholding eigenvalues of the Hessian matrix at each pixel to detect elongated structures. While computationally efficient, these methods may struggle to recover faint or low-contrast filaments across a large dynamic range of CD. Global approaches, including multi-scale decomposition techniques such as \getsf, aim to recover the spatial extent of filaments more comprehensively but can be computationally intensive when applied to large surveys. Beyond these practical limitations, there are inherent differences in the definition of filaments, leading to inconsistent results \citep{Hacar_2023}. This highlights the need for a more unbiased and parameter-free algorithm for filament identification and characterization.

    More recently, convolutional neural networks, particularly U-Net architectures, have been applied to Galactic filament detection. \cite{unetfilZavagno2023} presented a supervised U-Net/U-Net++ segmentation trained on filaments extracted by \cite{2020MNRAS.492.5420S}. \citet{berthlot} introduced position encoding to improve performance across large Galactic-plane regions with varying density and contrast. However, these approaches typically rely on threshold-derived filament masks for supervision and treat segmentation as a purely image-based task, separate from downstream physical characterization.

    In parallel, dedicated tools such as \texttt{FilFinder} and \texttt{RadFil} have established strong pipelines for filament skeletonisation and radial profile extraction. These tools have been widely adopted for deriving filament widths and Plummer profile parameters. Nevertheless, identification and physical characterization are generally performed sequentially. Filament characterization reveals multiple properties which can act as metrics to identify realistic filament profiles out of the extracted skeletons. Therefore, the filament identification problem can benefit from inputs given by filament characterization, highlighting the need for an integrated approach.

    In this work, we introduce \sutra : a unified framework for parameter-free, computationally efficient filament detection and characterization. The name \sutra is taken from {\em Sanskrit} for `thread', aptly representing the purpose of this tool, to unravel the threads (the filaments) of the ISM.\footnote{ Similar to the source identification tool \textsc{cutex} \citep{cutex}, \sutra will be made available \tcm{through} an online portal https://github.com/KumaranShivam5/sutra}. \sutra aims to combine the merits of different identification algorithms, as well as constraints obtained by characterization routines, to provide a machine-learning based integrated filament identification and characterization framework.

    \sutra combines four tasks into a common filament identification and characterization framework. These include (1) training segmentation models on the union of complementary identification strategies, (2) utilisation of filament skeletons instead of threshold-based masks as training labels, (3) incorporating beam-level physical characterization as a filtering criterion for candidate skeleton segments and (4) providing a modular, parameter-light framework that enables end-to-end automation across surveys. Together, these tasks enhance the robustness of \sutra across regions of varying contrast and enable direct learning of filament crests. By coupling segmentation with physical plausibility constraints, \sutra transforms filament extraction from a purely image-based problem into a physically guided procedure. This integration enables robust identification while simultaneously producing beam-resolution maps of filament physical properties, facilitating studies of fragmentation, stability, and star formation.
    
    In presenting this study, the paper is structured as follows. The data used in the study is described in \S \ref{sec:data}. The methodology used for identification of the filament is given in \S\ref{sec:identification-methodology}. The characterization methodology is discussed in \S\ref{sec:char}. The validation and comparative study of \sutra is given in \S\ref{sec:res-dis} section. The summary of the work, implications of \sutra, its limitations and future scope is discussed in \S\ref{sec:conclusion} .

\section{Data used}\label{sec:data}
   
     Observations from \textit{Herschel} represent a large sample for studying filaments in interstellar medium. The \textit{Herschel} telescope carries two photometric instruments namely Photodetector Array Camera and Spectrometer (PACS) and  Spectral and Photometric Imaging Receiver (SPIRE). The PACS instrument \citep{PACS} covers the wavelength range $\sim 60 - 210\mu m$ and is capable of both photometric observations and imaging spectroscopy. The SPIRE \citep{SPIRE} instrument contains an imaging photometer operating at $250$, $350$ and $500\mu m$ and an imaging Fourier-transform spectrometer covering wavelength range of $194 - 671 \mu m$. In this paper, we use CD maps generated from PACS and SPIRE observations taken as part of the HGBS survey \citep{Andre_2010}. The HGBS survey is one of the key science surveys by \her and covers 15 star forming regions spanning $\sim 160 deg^2$ in the Gould Belt which hosts a bulk of nearby ($<500pc$) molecular clouds \citep{Konyves_2015}. The survey covers a wide range of environments from quiescent to active cluster forming regions. The CD maps are obtained from the HGBS Survey archive\footnote{\url{https://gouldbelt-herschel.cea.fr/archives}} \citep{arz2019} to trace the filaments. The CD maps used in this work have a spatial resolution corresponding to the lowest resolution band i.e. SPIRE $500 \mu m$ with Half Power Beam Width ($HPBW$) of $36.3$\arcsec (or $12~pixels$).

\section{Methodology for filament identification}\label{sec:identification-methodology}

    This section describes the end-to-end methodology of \sutra for filament crest detection and subsequent skeleton-based characterization. Our task is to train \sutra to learn filament crest likelihood directly from column density maps without explicit intensity or curvature thresholding. We first summarize the training-label construction (Sec.~\ref{sec:training-data}), then introduce the U-Net-based model (Sec.~\ref{sec:ridge-model}), the preprocessing and patching strategy (Sec.~\ref{sec:training-method}), the training and loss choices (Sec.~\ref{sec:model-training}), and finally the conversion of the model output to a binary skeleton and the physics-guided skeleton refinement (Sec.~\ref{sec:output-to-skel} and \ref{sec:physics-refine}).

\subsection{Training Data Construction: Skeleton map using \getsf and \disperse}\label{sec:training-data}

     We build training labels from the union of two complementary filament-finding methods: \disperse (crest/topological persistence) and \getsf (global multiscale extraction). \disperse preferentially traces intensity ridges while \getsf recovers extended filamentary extent across scales. The union of their outputs gives a consensus set of filament crests that is less biased by any single algorithm's assumptions and more inclusive of low-contrast structures.
    
    \begin{figure}
        \centering
        \includegraphics[width=0.5\textwidth]{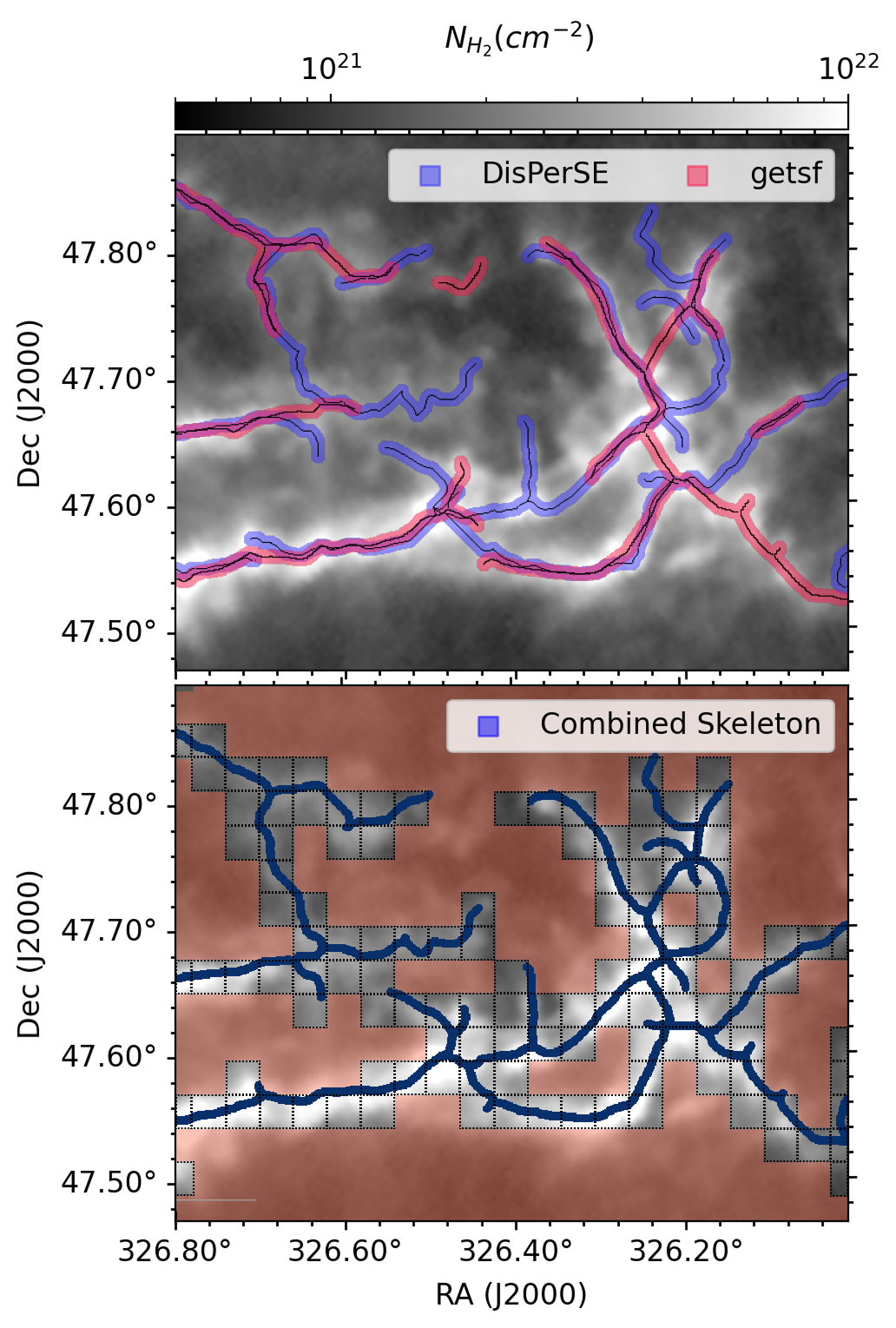}
        \caption{Comparison of filaments identified using \disperse and \getsf. The top panel compares filaments identified by \disperse (blue) and \getsf (red) in an IC5146 cloud cutout. The bottom panel shows the combined skeleton (\ufil skeleton) used for \unet training. Grayscale regions are training chunks for the ML model; red regions are excluded. }
        \label{fig:training-skeleton}
    \end{figure}

    Table \ref{tab:training-table} lists the molecular clouds, their distances, and pixel areas used. To create our training filament skeletons, we combine two widely-used identification tools: \getsf and \disperse. These methods differ significantly in their filament-definitions and extraction algorithms.

   \getsf smoothens the input CD map at reducing scales iteratively, identifying filaments as the structures that maintain their aspect ratio across these scales. It's largely parameter-free, though computationally expensive, often taking days (e.g., 3-4 days see Table \ref{tab:training-table} column-4). \disperse traces filaments as integral lines connecting saddle points to minima. It is sensitive to two parameters, `persistence' and `robustness' thresholds, which must be globally selected. The use of global thresholds limits the algorithm as it can miss structures in clouds with high CD variations \citep{arz2019} and also in low density regions in the CD map. As Figure \ref{fig:training-skeleton} (top panel) shows, while \getsf and \disperse often overlap, each also identifies unique filamentary regions. The differing total filament lengths in Table \ref{tab:training-table} further underscore these discrepancies.

      \begin{table*}[t!]
      \centering

         \centering
         \small
             \begin{tabular}{lccccccr} 
                \toprule \toprule
                &  & & \multicolumn{2}{l}{\textbf{\getsf}}  & \multicolumn{3}{l}{\textbf{\disperse}}   \\
                Field & Size & d & Time & L & PT & RT & L \\
                & ($pixels$) & ($pc$) & (hrs) & ($pc$) & \multicolumn{2}{l}{($ 10^{21}cm^{-2}$)}  & $(pc)$ \\
                (1) & (2) & (3) & (4) & (5) & (6) & (7) & (8) \\
                
                \midrule
                \textbf{Ophichus} & 5.7 $\times$ 6.0 & 140 & 99.8 & 33 & 0.01 & 1.48 & 24 \\
                \textbf{Polaris} & 6.5 $\times$ 6.2 & 150 & 83.2 & 22.6 & 0.17 & 0.84 & 5.2 \\
                \textbf{Aquila} & 5.2 $\times$ 5.6 & 260 & 81.9 & 45.2 & 0.2 & 4.06 & 34.1 \\
                \textbf{Orion} & 10.3 $\times$ 8.1 & 400 & 82.5 & 65.2 & 0.17 & 1.09 & 186.4 \\
                \textbf{IC-5146} & 3.8 $\times$ 3.0 & 460 & 112.6 & 33 & 0.17 & 1.0 & 33.5 \\
                \bottomrule
            \end{tabular}

             \caption{ Cloud complexes used for training \unet model. Columns: (1): Cloud name (2): Footprint area of CD maps in $1000\times 1000 pixels$ unit; (3): Cloud Distance(4):CPU wall-time taken by \getsf; (5): Total length of \getsf filaments; (6) and (7): \disperse parameters: persistence and robustness, respectively\citep{arz2019}; (8): Total length of \disperse filaments}
        \label{tab:training-table}
    \end{table*}

    With \getsf and \disperse we extract skeleton map for CD map for five nearby molecular clouds CD maps: Ophiuchus, Polaris, Aquila, Orion-B (referred to as Orion hereafter) and IC-5146 at distances, $140pc$, $150pc$, $260pc$, $400pc$ and $460pc$, respectively. Table \ref{tab:training-table} lists the molecular clouds, their distances, and pixel areas used. In \getsf, we set the {\em maximum filament length} parameter to $180\times HPBW$ for all fields. In \disperse, we use {\em persistence} and {\em robustness} threshold (columns 6 and 7 in table \ref{tab:training-table}) derived for these clouds in \cite{arz2019}. 
    
    We generate single-pixel-wide union skeleton using morphological transformations with \textsc{scipy} package \citep{2020SciPy-NMeth}. This involves widening individual skeletons, creating a union map of these widened filaments, and then re-skeletonizing the result via medial axis transform. Finally, small structures (Length $<3\times HPBW$) are removed. The top panel of Figure \ref{fig:training-skeleton} shows the dilated skeleton obtained from \disperse and \getsf. The single pixel wide skeleton(black curves in top panel) do not exactly overlap, before dilation. The final combined skeleton map, hereafter referred to as \textbf{\ufil} is shown in the bottom panel of Figure \ref{fig:training-skeleton}. Training chunks are sampled from five HGBS fields (Table \ref{tab:training-table}); chunks without any skeleton pixels are excluded from training to focus the model on crest localization.

\subsection{Filament-Ridge detection pipeline : model objective and architecture}\label{sec:ridge-model}
    \begin{figure*}
        \centering
        \includegraphics[width=0.99\linewidth]{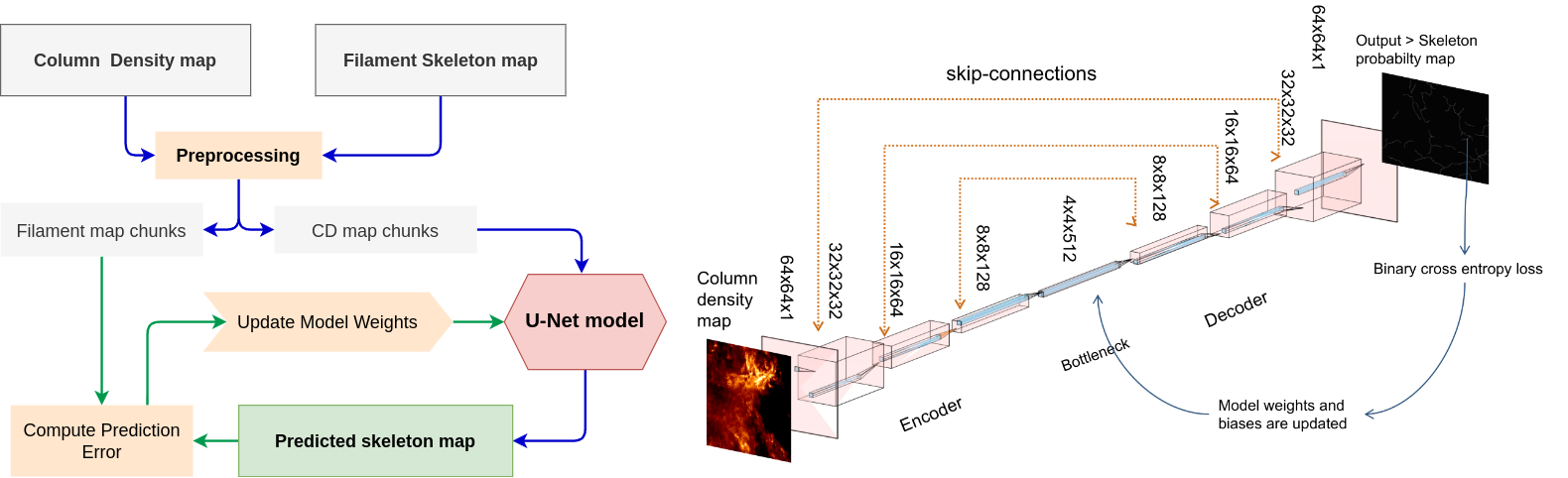}
        \caption{
        Filament Identification workflow (left). Inputs for the algorithm is CD map and the skeleton map, which is passed through the preprocessing module \S\ref{sec:training-method} followed by input to the ML algorithm. Architecture of \unet model with image subdivision size $64\times 64$ (\S \ref{sec:training-method}). The size of each layer is indicated in the figure. The dotted orange lines represent the skip-connections between the encoder and decoder layers.
        }
        \label{fig:u-net}
    \end{figure*}

    Rather than learning a binary filament region, the U-Net model is trained to produce a \emph{crest-likelihood} (ridge-probability) map: a continuous-valued image $P_{\rm crest}(x,y)\in[0,1]$ that indicates the probability that a pixel lies on a filament crest. Training targets are single-pixel skeletons, so the network optimizes ridge localization (centerline detection) instead of interior/edge segmentation. This distinction ensures that network predictions concentrate probability along narrow crests and that subsequent thresholding selects high-confidence ridge pixels rather than defining filament geometry.
    
    We evaluate several U-Net variants denoted UN$K$ (e.g., UN64 for $64\times64$ input chunks). The encoder–decoder backbone follows the standard U-Net design with skip connections; the final layer uses a sigmoid activation to produce $P_{\rm crest}$. Figure \ref{fig:u-net} (right panel) shows the adopted architecture for UN64.

    \subsubsection{Preprocessing, patching, and overlap strategy}\label{sec:training-method}

    \unet models require a fixed input image size, but our CD maps vary greatly in pixel dimensions (Table \ref{tab:training-table}). To address this, we divide the large CD maps into smaller `chunks' that match the \unet's specific input size. These chunks undergo the following preprocessing before U-Net training:
    \begin{itemize}
        \item Global background filtering: The histogram of CD values are divided into bin size of $\nh = 10/cm^{2}$. The median of first bin is taken as the global background. The pixels where CD value is less than this global threshold is set to zero.
        \item Subdivision: We divide both the CD map and its skeleton map into non-overlapping chunks. For training, we discard WCS information (it's retained in the \sutra pipeline for later rejoining). Chunks lacking any skeleton pixels are removed.
        \item Normalization: A local normalization of the individual subdivided CD map chunks ($CD^{i}$) is carried out, and is given by:
        \begin{align}
            CD^i_{norm} = \frac{CD^i-mean(CD^i)}{\sigma(CD^i)^2}
        \end{align} 
        Here, $CD^i$ denotes the $i^{th}$ chunk, and $\sigma$ is its CD standard deviation. This local normalization is crucial for model robustness, preventing bias towards high-intensity regions. In \sutra's application, overlapping chunks mitigate variance from this normalization, ensuring more connected filaments.
        \item Contrast enhancement: To improve the contrast across all the chunks we apply chunk level flattening operation to the normalized chunks obtained from previous step:
        \begin{align}
            CD^i_f  = f \times tan^{-1}\left(\frac{CD^i_{norm}}{f}\right)
        \end{align} 
        where, $f$ is the flattening threshold, calculated as the $95th$ percentile of the normalized $CD$ value's distribution.
    \end{itemize}
    Chunk level normalisation can lead to inconsistencies between neighbouring chunks. To mitigate this, during inference we use a high-overlap tiling (95\% overlap for 64$\times$64 chunks, i.e., stride=4 px) and average the predicted crest-probabilities in overlap regions. This results in total 225 patches contributing to the output at each pixels. Averaging removes boundary discontinuities introduced by per-chunk normalization and yields a smooth full-map $P_{\rm crest}(x,y)$.

    \subsubsection{The U-Net model}\label{sec:the-model}
    We experiment with various derivatives of the \unet model for filament extraction. Depending on the subdivision size ($K$), various architectures of \unet, referred to as UN$<$K$>$, are examined in this work. As shown in Figure \ref{fig:u-net} (right panel), the \unet has an encoder, bottleneck, and decoder. Each encoder block uses $3\times 3$ convolution and max-pooling, halving the image size progressively until the bottleneck forms $1\times 1 \times N$ features (N being total input pixels).

    Decoder blocks use upsampling and convolution, doubling spatial resolution and halving feature depth. Skip-connections (dashed lines in Figure \ref{fig:u-net}) are vital for preserving input information. The final decoder output matches the input CD map size. We use rectified linear unit (ReLU) activation in layers and \textsc{sigmoid} activation function($\sigma(x) = 1/(1+e^{-x})$) at the output to map CD values to trace the crest of filament spine. The U-net model effectively acts as the following mapping function ($f$) with learnt parameters $\theta$:

    \begin{equation}
        f_\theta : \Sigma_{CD} \rightarrow P_{crest}(x,y) ,
    \end{equation}
    where $\Sigma_{CD}$ is the input CD map and output of the model is probability of a pixel $(x,y)$ to belong to a filament crest.

    \subsubsection{Training}\label{sec:model-training}
    
    We design and compare four \unet model architectures: UN32, UN64, UN128 and UN256 for chunk of sizes $32\times 32$, $64\times 64$, $128\times 128$ and $256\times 256 \,pixels$. Our goal is pixel-wise binary classification, generating one-pixel-wide filament skeleton map. However, due to the observation beam size, a perfect pixel-to-pixel alignment with the \ufil skeleton (\S\ref{sec:training-data}) is not possible. To address this, we dilate the \ufil skeleton map to $HPBW$ size before training. The loss function is then calculated between the model's predicted filament probability map and the dilated skeleton map for each subdivision. For training, Adam optimizer with learning rate $10^{-4}$ is used for minimizing loss function. Since segmentation is binary classification with very high class imbalance (very few skeleton pixels compared to background pixels), we use weighted binary cross entropy loss, defined as:
   \begin{equation}
        \begin{split}
            \mathcal{L}_{WBCE} = - \frac{1}{M} \sum_{i=1}^{K}  \sum_{j=1}^{K} \left[ \right. & w_1 y_{ij} \log(\hat{y}_{ij}) \\
            & \left. + w_0(1 - y_{ij}) \log(1 - \hat{y}_{ij}) \right]
        \end{split}
        \label{eq:wbce}
    \end{equation}
    where $M$ is the number of samples in the training batch, $K$ is the CD chunk size, $\hat{y}_{ij}\in[0,1]$ is the predicted crest probability $P_{crest}(i,j)$ and $y_i\in\{0,1\}$ the skeleton label for the $(i,j)$ pixel. We assign weights: 1 for skeleton pixels($\hat y_i=1$) and 0.1 for background pixels($\hat y_i=0$). These weights are determined through non-exhaustive hyper-parameter tuning. For segmentation task with highly imbalanced class distribution, DICE Loss (DL)\citep{diceloss2016arXiv160604797M} is another common alternative, defined as:
    \begin{align}
        \mathcal{L}_{Dice} = 1 - \frac{2 \sum_{i,j=1}^{N} y_{ij} \hat{y}_{ij} + \epsilon}{\sum_{i,j=1}^{N} y_{ij}^2 + \sum_{i,j=1}^{N} \hat{y}_{ij}^2 + \epsilon}\label{eq:dice}
    \end{align}

    For each selected UN$<$K$>$ model, the CD and \ufil skeleton maps are re-projected to a common pixel grid using \textsc{reproject} package. These maps are then divided into K-sized chunks. Chunks without skeleton are discarded. This dataset is randomly shuffled. 80\% of the chunks form the training dataset while the remaining 20\% are used as unseen test dataset. For training, non-overlapping chunks are generated to ensure no information leakage in the test-data. For each \unet architecture (UN32, UN64, UN128, UN256), we experiment with both the WBCE and DL loss functions. After each training epoch, we evaluate performance on test-data using the intersection over union metric(IoU), defined as:
    \begin{align}
        IoU = \frac{\text{Area of Overlap}}{\text{Area of Union}} = \frac{\hat Y \cap Y}{\hat Y \cup Y}
    \end{align}
    \begin{figure}[ht!]
        \centering
       \includegraphics[width=0.98\linewidth]{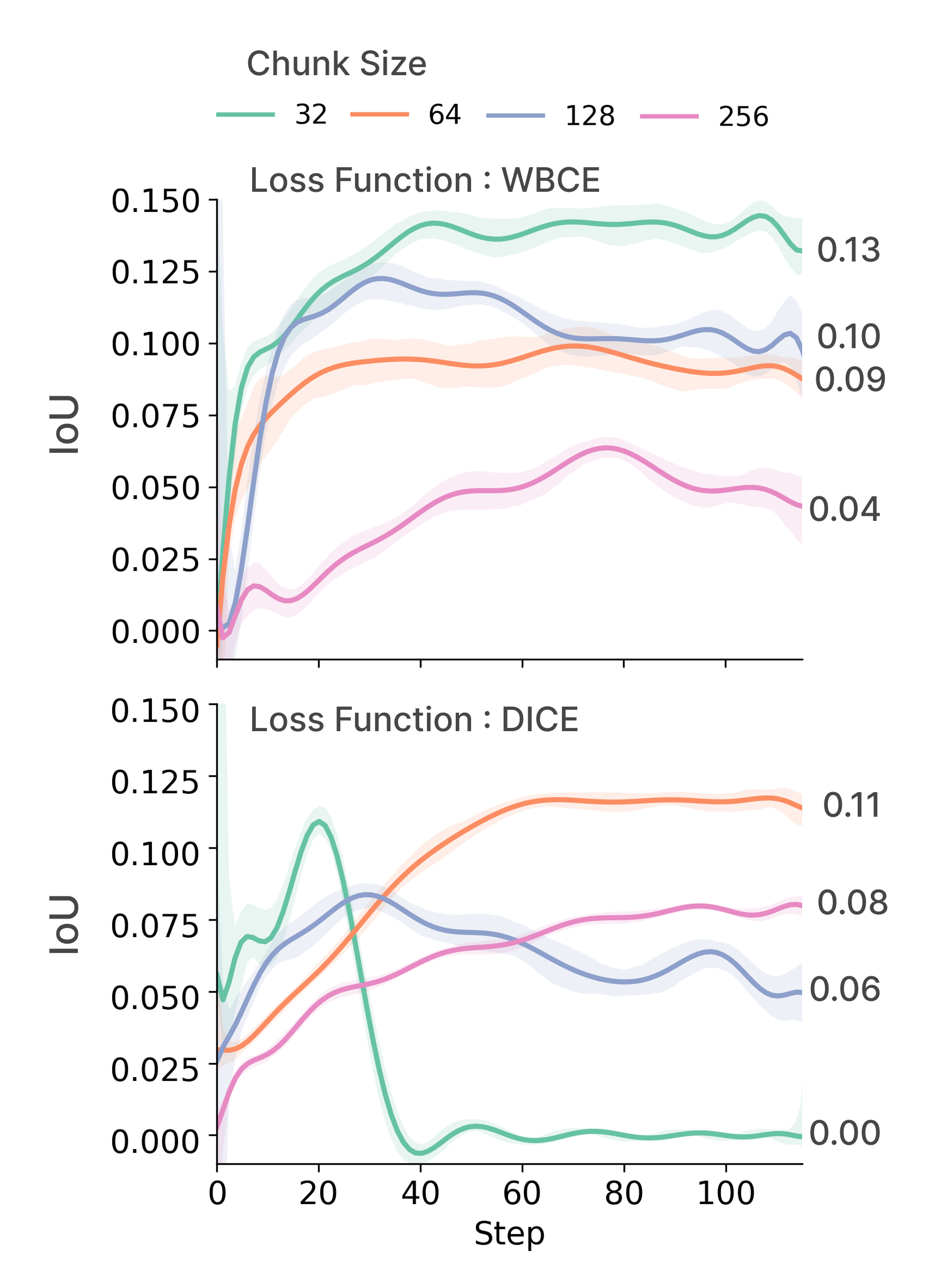}
        \caption{The loss on the validation dataset with training epoch for different models considering different chunk sizes from 32 pixels to 256 pixels for loss functions : WBCE (top) and DICE loss (bottom)}
        \label{fig:training-loss}
    \end{figure}
    
    where $\hat Y$ is the chunk of \ufil skeleton in the test-dataset and $Y$ is its corresponding model prediction. The IoU on the test-data after each iteration is computed. This exercise is done for combination of all the four \unet architectures and the loss functions. The IoU at each training epochs is shows in the Figure \ref{fig:training-loss}. In each iteration, the model is trained till IoU reach convergence. For WBCE, the smallest chunk model UN32 performs the best, but for DL, after 20 steps, the model moves out of minima and fails to converge. DICE loss gradient is inherently unstable. A small receptive field result in minimal overlap between predicted and training skeleton, which results in steep jump in DICE loss due to very small value in denominator (eq.\ref{eq:dice}) \citep{dicelossYEUNG2022102026}. WBCE loss gradients are smooth as loss for every single pixel independently (eq.\ref{eq:wbce}).. UN64 performs best for DICE loss and has similar IoU for BCE loss. For both losses, IoU achieved by UN64 and UN128 models are very similar (0.09 and 0.1 respectively). Moreover, the UN64 model surpasses UN128 model for DICE loss. Considering the overall higher IoU for UN64 model and comparatively better resilience across loss function, UN64 is selected for further analysis. 

    \textbf{Resolution and patch-size considerations.} The chosen patch size (e.g., UN64) and the survey pixel scale set the angular range of features the network most naturally represents approximately up to half the patch width. For more distant surveys (e.g., Hi-GAL) where filaments subtend few pixels, two adaptation paths are available: (i) reproject the input maps to match the training pixel scale (flux-conserving reprojection), or (ii) apply transfer learning by fine-tuning the final network layers on resampled data. Appendix \ref{app:pixel-scale} demonstrates a controlled test applying the trained model to Hi-GAL maps at native and resampled scales. \tcm{Appendix \ref{app:pixel-scale} also discusses the impact of resolution degradation and downsampling, treated as an effective distance scaling, on \sutra using a high-resolution Aquila HGBS cutout.}

    \subsection{Model Output to Skeleton}\label{sec:output-to-skel}
    
     \begin{figure}
        \centering
        \includegraphics[width=0.48\textwidth]{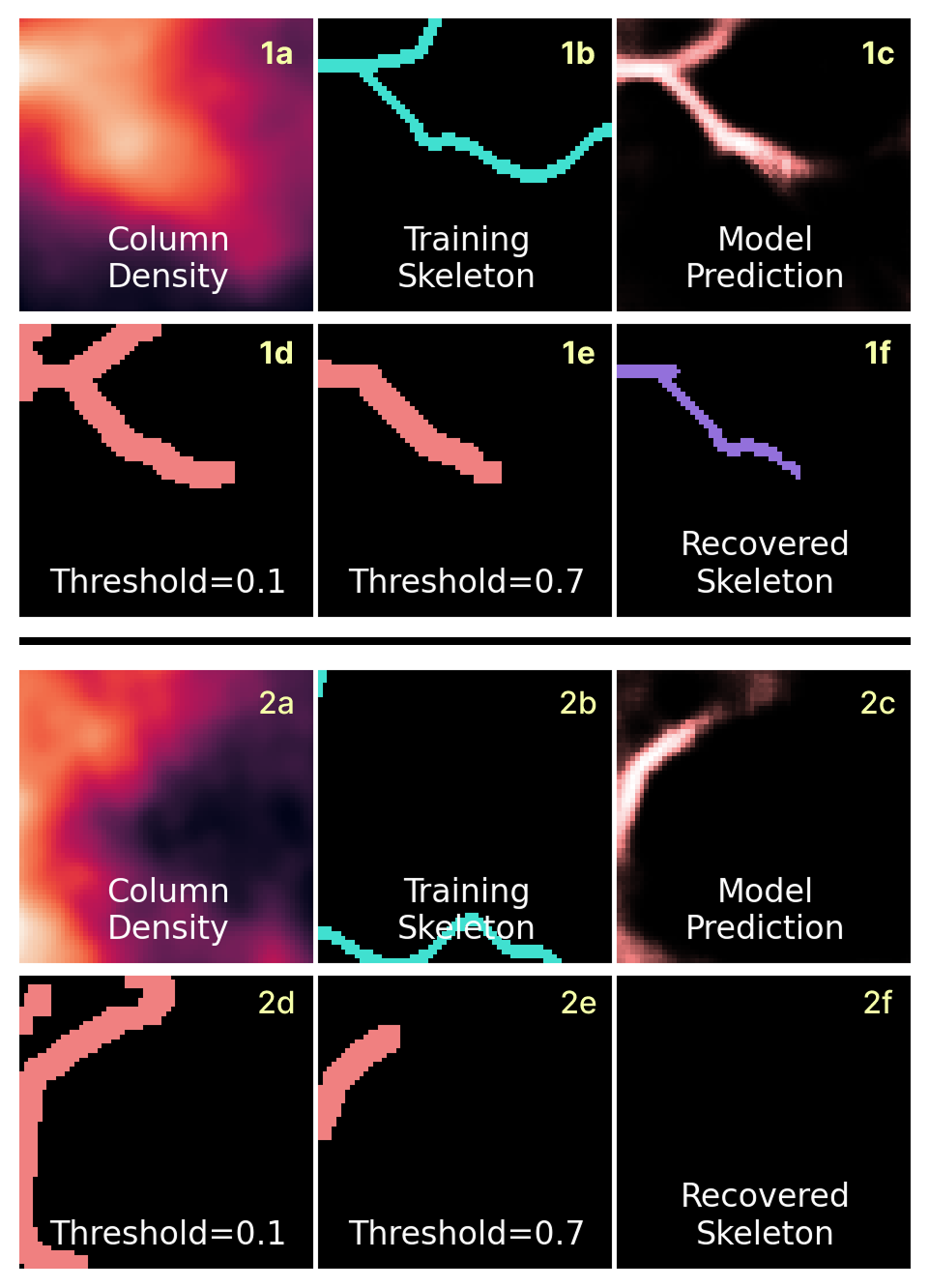}
        \caption{Illustrations for algorithm adopted for evaluation of recovering capability of the \unet model in \sutra. The panels are (a): Chunk of CD map; (b): \ufil skeleton corresponding to this chunk (\S\ref{sec:training-data}); (c): output of the \unet model with values lying between 0-1; (d) and (e) : the model output is converted to threshold mask (0.1 for (d) and 0.7 for (e) panel), skeletonised and then dilated to the beam size of 12 pixels; (f): the intersection of 0.7 threshold skeleton (panel (e)) and the \ufil skeleton (panel (b)).}
        \label{fig:skeleton-compare}
    \end{figure}
     \begin{figure}
        \centering
        \includegraphics[width=0.48\textwidth]{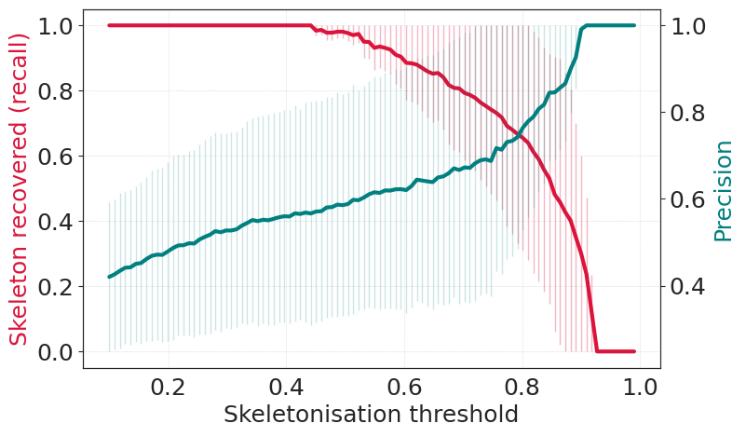}
        \caption{Fraction of filament in the test-dataset recovered by the \unet model after thresholding and skeletonization. The solid red curve shows the median of the relation over all the chunks in test-dataset and the vertical errorbars show median absolute deviation. The solid blue line shoes the fraction of predicted skeleton which also overlaps with the \ufil skeleton.} 
        \label{fig:recovery-curve}
    \end{figure}

    The trained model produces a full-map crest-likelihood $P_{\rm crest}(x,y)$. To obtain a binary crest map we apply a threshold $S_t$ to select high-confidence ridge pixels, then perform a medial-axis transform to ensure single-pixel connectivity. Crucially, the role of $S_t$ in \sutra differs from that in threshold-based approaches where the threshold defines a broad filament mask whose topology determines the skeleton; in \sutra thresholding selects pixels from an already narrow (see Figure \ref{fig:Plummer-example} top-middle panel), crest-focused probability map, so $S_t$ controls confidence rather than geometry.

    To quantify the recovered skeleton fraction we perform the following steps:
    \begin{itemize}
        \item Convert the UN64 model output $P_{crest}(x,y)$ into binary skeleton mask by applying a threshold value ($S_t$)
        \item Skeletonize this mask using medial axis transform\citep{zhang},
        \item Dilate the skeleton by half the beam-size ($HPBW/2$ \tcm{or} $6 pixels$) on each side.
    \end{itemize}
     Figure \ref{fig:skeleton-compare} illustrates this: panels (a) and (b) show the CD map and \ufil skeleton; (c) displays the U-Net output. Panels (d) and (e) show thresholded, skeletonised, and dilated results for thresholds $S_t=0.1$ and $S_t=0.7$, respectively. The recovered skeleton is then determined by intersecting the dilated skeleton with the training skeleton (e.g., panel 1e compared to 1b in Figure \ref{fig:skeleton-compare}).

    We examine the recovery of the \ufil skeleton as a function of the skeletonization threshold $S_t$ across all test chunks (Figure~\ref{fig:recovery-curve}). The median recovery curve (red line) shows the fraction of \ufil skeleton pixels recovered by the model for different values of $S_t$. We also compute the precision, defined as the fraction of predicted skeleton pixels that overlap with the \ufil skeleton relative to the total predicted skeleton length. For $S_t > 0.8$, more than 70\% of the predicted skeleton overlaps with \ufil, while the remaining fraction can be interpreted as potential spurious detection.

    At the standard classification threshold $S_t = 0.5$, the U-Net recovers more than 98\% of the \ufil skeleton. As $S_t$ increases, recovery decreases, with 80\% of the \ufil skeleton still recovered for $S_t < 0.7$. Visual inspection of chunks with recovery below 80\% frequently reveals incomplete or mislabeled structures in the \ufil map (e.g., panel~2 of Figure~\ref{fig:skeleton-compare}). This panel also shows elongated ridge-like structures that are not traced by either \disperse or \getsf.

    The features marked as missing filaments correspond to high-aspect-ratio ridges that satisfy the morphological and contrast criteria commonly adopted in \textit{Herschel} studies \citep{Andre_2010,Arzoumanian_2011,arz2019}. These structures may therefore represent genuine filamentary features that are absent in the \ufil skeleton rather than spurious predictions. In the following section, we introduce the physical filtering procedure used to assess whether such structures are physically consistent filaments. We note that distinctions between filaments, fibers, and sheet-like structures can be ambiguous; throughout this work, we adopt the conventional \textit{Herschel}-based definition.

    A qualitative comparison of \sutra with other existing filament identification methods under controlled synthetic conditions is presented in Appendix \ref{app:comparison}. This comparison is intended to illustrate robustness under varying background levels and does not influence the training procedure or quantitative validation.
    
    \subsection{Physical filtering for skeleton refinement}
    \label{sec:physics-refine}

     \begin{figure*}
        \centering
        \includegraphics[width=0.9\textwidth]{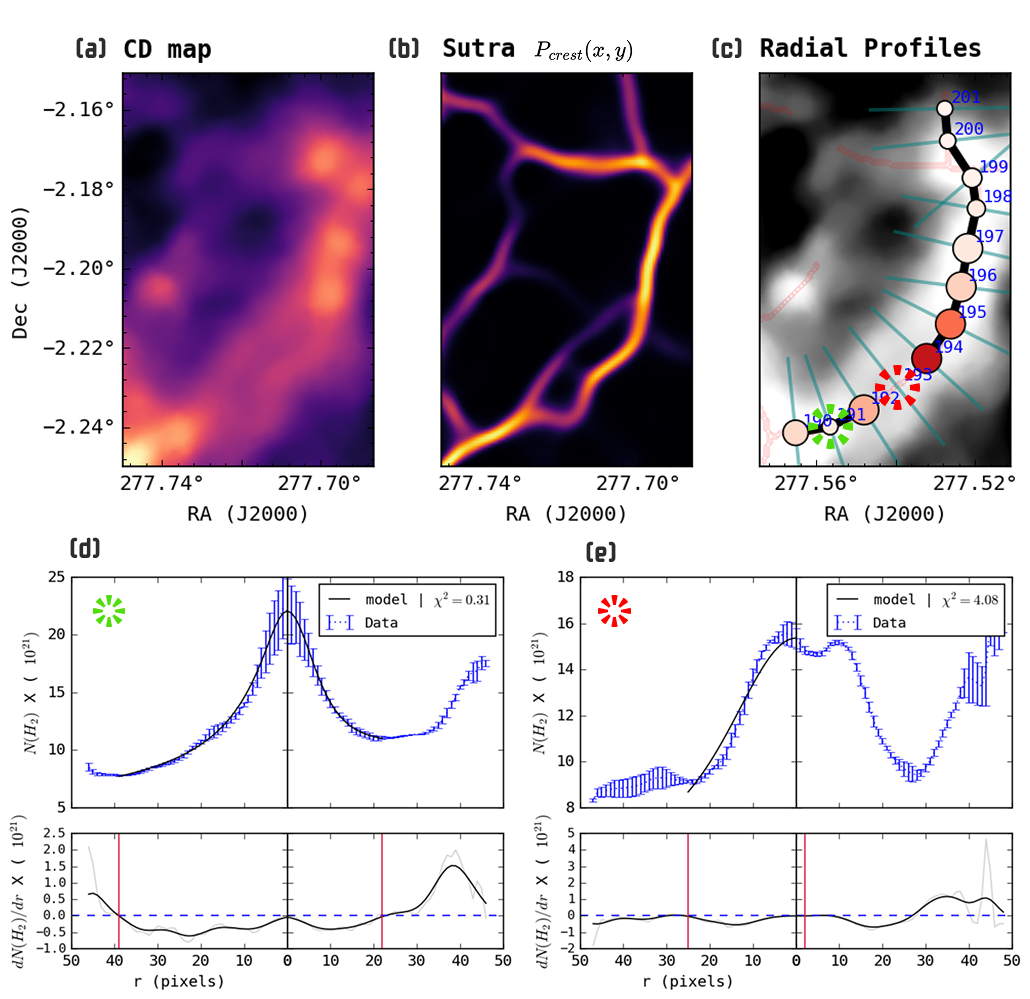}
        \caption{ Schematic illustration of crest detection and physics-guided skeleton refinement in \sutra (a) Input column density (CD) map. (b) Corresponding crest-likelihood map $P_{\rm crest}(x,y)$ produced by the trained U-Net ridge-detection model, where high-probability values trace filament spines. (c) Binary skeleton obtained after thresholding the crest-likelihood map and medial-axis refinement. Beam-sized segments (numbered) are evaluated individually through radial profile analysis. The bottom panels shows example of (d) 'good' beam segment (green marker) yielding a low low reduced $\chi^2$ and (e) shows example of rejected beam (red marker), due to poor Plummer fit, on left size and immediate dip ($\leq$ beam size) due to nearby filament on right side. The solid black line (panel c) shows the final skeleton.}
        \label{fig:Plummer-example}
    \end{figure*}
    
    First, the UN64 ridge-detection model predicts a narrow crest-likelihood map from the CD image, which is thresholded to obtain a binary skeleton. This skeleton is divided into beam-sized segments, and each segment is independently evaluated through radial profile fitting. Filaments from observation \citep{Ostriker_1964, andre2010initial, Arzoumanian_2011,Hacar_2023, arz2019, Hacar_2023} and simulations \citep{Federrath} shows cylindrical structure whose radial profiles are well represented by Plummer profiles. Here the quality of the Plummer fit and the filament contrast are used as physical diagnostics to determine whether the segment is consistent with a cylindrical filament structure.
    
    Segments that do not satisfy these criteria are rejected, while the validated beam elements are reconnected using a minimum spanning tree (MST) algorithm implemented in \texttt{scikit-learn} \citep{scikit-learn}. MST-based reconstruction has previously been used to connect dense structures into filamentary networks (e.g., \citealt{sedigism2023A&A...675A.119G}). In \sutra, this step ensures continuity only among physically consistent segments.
    
    This feedback-driven refinement reduces reliance on purely image-based thresholds and yields a filament catalog that is both morphologically consistent and has physical significance. Figure~\ref{fig:Plummer-example} illustrates this process. Examples of rejected beam segments are presented in Appendix~\ref{app:Plummer-filtering}. Details of the radial profile extraction and fitting procedure are described in the following section.

\section{characterization of identified filaments}\label{sec:char}

    Given the CD map, single-pixel-wide filament skeleton crest map is generated using the UN64 model following \S\ref{sec:identification-methodology}. In this section, we describe the methodology employed in \sutra to compute the physical properties of filaments.

\subsection{Radial Profile Extraction}\label{sec:rad-prof}

   \begin{figure*}
        \centering
        \includegraphics[width=1\linewidth]{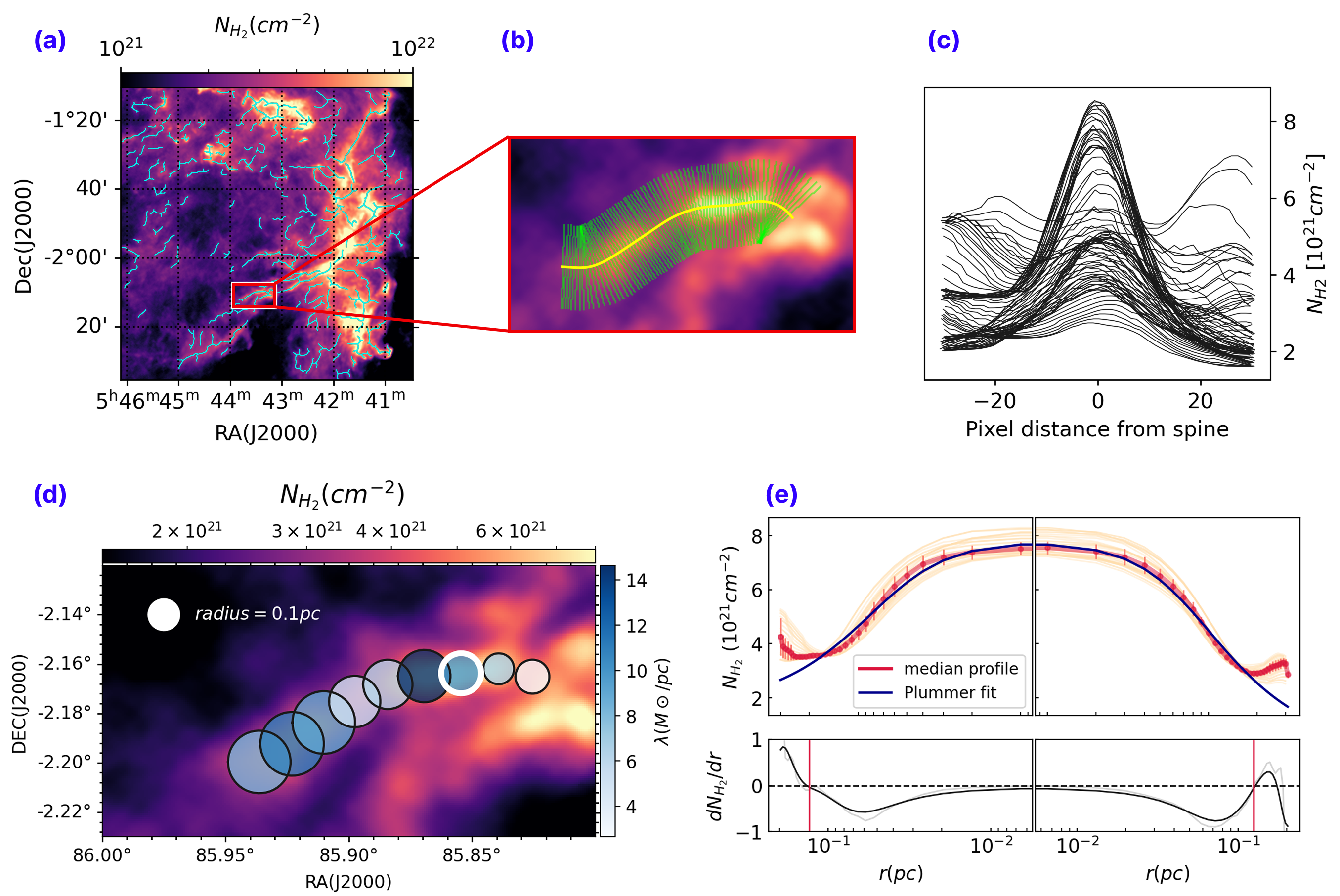}
        \caption{Illustration of workflow for local characterization of individual filament starting from the output of \unet identifier after skeletonization: (a) Skeleton obtained from the \unet identifier. (b) Zooming in to one filament, with the smoothened skeleton(shown in yellow curve) and lines along which radial profile is extracted (green lines). (c) radial profiles for the selected filament. (d) The radial profiles are grouped together at a distance of $HPBW$. Size of the circle shows local width ($W_{bg} = R_{bg}^++R_{bg}^-$) (e) Plummer profile fit to the the median radial profile for all the profiles within the group highlighted by white circle in (d). 
        }
        \label{fig:orion-example}
    \end{figure*}
 
    We adopt a crest-based radial profile methodology similar in spirit to RadFil \citep{radfil}, where profiles are constructed perpendicular to the filament skeleton. We first segment the skeleton map into individual, unbranched filaments.  Starting from an initial pixel, we iteratively find the next nearest pixel until all skeleton pixels are processed. Then, a distance threshold is applied such that when the distance between adjacent pixels exceeds this threshold, the filament is separated into different skeleton. This process effectively isolates each filament, and in intersection regions, automatically separates all but one branch and then proceeds for labeling the next skeleton. Separation of skeleton is necessary as, at regions close to the intersection between multiple filaments, the local tangents cannot be calculated accurately.
    
    In the next step, we compute local tangents to the filament skeleton. The skeleton map represents filaments as single-pixel-wide curves, despite their finite width in the CD maps. This representation, combined with the beam width of the CD map ($HPBW = 36.3\arcsec$ or $12pixels$) causes variations and abrupt changes in the calculated tangent directions along the skeleton. These changes can lead to inaccurate radial profiles. To address this and align the skeleton curve with the filament axis, we smooth the curve using a B-spline fit\footnote{\disperse performs similar skeleton smoothing using \texttt{skelconv} algorithm}. The smoothing parameter (max-knots of B-spline) is set to one-third of the beam width, effectively reducing pixel-level noise within the beam's influence. Using this smoothened skeleton, we select a kernel of size 5 pixels around a local region of the skeleton, and fit a line along the pixels. After obtaining the slope($m$) of this fitted line, we calculate the slope of the line perpendicular to this line ($-1/m$).
    
    An example of this methodology is shown in Figure \ref{fig:orion-example}. Panel (a) shows a cutout of the Orion cloud complex and the identified filaments skeleton map using UN64 model overlaid on top of its CD map. For demonstration, a single filament is highlighted from this cutout, shown in panel (b). The local tangents are calculated with a kernel of 5 pixels and a 2 pixel stride. The yellow curve shows the filament skeleton and the lines perpendicular to the filaments are shown in green. We choose a maximum pixel distance (30 pixels) that is traversed along this perpendicular line to obtain CD values as the local radial profile. In some cases, it may be required to obtain the radial profile at larger distances (e.g when analyzing very isolated filament like Musca \citep{musca}). This max radial profile distance is given as an adjustable parameter in \sutra, giving the user an option to obtain filament profile till the region of interest.

    The local radial profiles are not necessarily centered on zero due to smoothing of the skeleton. Therefore, we shift each profile so its maximum value aligns with r = 0. The environments on either side of the filament differ, (significantly in some cases where another filament lies in the vicinity on one side, or at the places where filament has very high curvature) so each radial profile is split in half at the center for subsequent analysis and local properties are computed on each side separately.

\subsection{Filament physical properties}\label{sec:phy-prop}
    
    Typically filament properties are computed by taking the median of all profiles. However, a filament's physical properties can vary significantly along its length, meaning a single filament-level median might not accurately represent its true bulk characteristics \citep{howard}. In \sutra, we address this by computing filament properties at the beam-size resolution. We do this by subdividing each filament into smaller segments of the size of $HPBW$ and grouping together all local radial profiles within this segment.\footnote{in \sutra user can control this subdivision length $\Delta l$}. Panel (d) in Figure \ref{fig:orion-example} illustrates this beam-level grouping. The circles represent filament-segments with $1\times HPBW$. Median of grouped-radial profile (and individual local profiles) is shown in panel (e) (as light red curves) for the profile group highlighted in panel (d). The dispersion in median profile is given by median absolute deviation, computed as: 
    \begin{equation}
        \Delta \nh(r) = \text{median}\left[|\nh^i(r) - \text{median}(\nh(r))|\right]\label{eq:del_nh}
    \end{equation}
   where $\nh^i(r)$ represents the $i^{th}$ profile in the given segment. This grouping via median of local radial profiles is used to compute the local beam-level filament properties. The filament bulk properties are drawn from the distribution of beam-level statistics.

    Finding the radial extent of a filament is a crucial step for subsequent characterization. In crowded or branching environments, manual masking is often required to isolate overlapping structures prior to profile fitting. In \textsc{RadFil} this masking is done with the help of \textsc{pts\_mask} argument. Otherwise the parameter \textsc{fildist} can be set manually to restrict fitting upto certain distance on either sides. To enable fully automated processing, in \sutra we adopt a truncation criterion inspired by \citet{arz2019}, where the radial extent is defined by the point at which the density gradient flattens toward the background.
    
    Due to their complex hierarchical nature, filaments are rarely seen in isolation \citep{Hacar_2023}. In frequent cases the radial profile of the filament may overlap with other nearby filaments (see panel (b) \ref{fig:orion-example}). This results in a rise in radial profile after the first minima. The radial distance upto which fitting is performed should not include contribution from these nearby filaments. To determine the outer radius, we find out the slope of filament radial profile $d\nh(r)/d{r}$. Next, the slope is convolved with a Gaussian kernel with a standard deviation of $2.3\times (HPBW)$ (see panel (e) of Figure \ref{fig:orion-example} and panel (d) of Figure \ref{fig:Plummer-example} where right side profile is truncated at ~20 pixels). The outer radius $R_{bg}$ is the distance at which the filament merges with the background, thus having zero slope: 
    
    \begin{equation}
        \frac{d\nh(r)}{dr}_{\left(r = R_{bg}\right)} = 0
    \end{equation}

    The radius $R_{bg}$ is computed seperately on each side of the filament axis (referred as $R_{bg}^+, R_{bg}^-$).
    To compute the physical properties we follow the assumption that filaments are linear structures, following cylindrical profiles \citep{arz2019, hacar2023ASPC..534..153H}. The 2-dimensional projection of CD follows Plummer-like radial profile \citep{ostriker1964ApJ...140.1056O} with the CD value ($\nh(r)$) at a distance $r$, perpendicular to the filament axis is given by:
    \begin{align}
        \nh(r) = \nh^{bg} + \frac{\nh^{0}}{(1+(r/R_{flat})^2)^{(p-1)/2}}
    \end{align}
    where $R_{flat}$, $p-index$ ($p$) and $\nh^{bg}$ are free parameters. $p-index$($p$) of Plummer fit holds significant physical implications. For hydrostatic filament in isothermal conditions, $p=4$ \citep{stodolkiewicz_1963,Ostriker_1964}. Filaments in the nearby cloud in HGBS survey mostly show much shallower fall-off with $p\sim2$ \citep{Kirk_2013, arz2019}. Plummer fit is done only till the outer radius $R_{bg}$. We perform the fit using \texttt{scipy's} \texttt{curve\_fit} routine \citep{2020SciPy-NMeth}, The goodness of fit is measured using reduced $\chi^2$ statistic. Error ($\sigma$) at each point for $\chi^2$ is taken as the $\Delta \nh(r)$ (eq. \ref{eq:del_nh}). The profiles which fail to fit are fitted again after freezing $p=2$. The profiles failing this fit are also flagged and dropped from statistics and the corresponding table row is populated with \textsc{NAN} values.

    Linear mass density of the $i^{th}$ filament segment is computed by integrating the radial CD profile $\nh(r)$ on both sides of the segments upto the outer radius point on each side represented by $+$ and $-$
    \begin{align}
        \lambda_i = \int_{r = 0}^{R_{bg}^+} \left(\nh^+(r) - \nh^{bg+}\right)dr + \nonumber \\ \int_{r = 0}^{R_{bg}^-} \left(\nh^-(r) - \nh^{bg-}\right)dr
    \end{align}
    
    Total mass is obtained by summing up the linear mass density $M = \sum_i \lambda \Delta l = \Delta l \times \sum_i \lambda$.

    The contrast of filament is also defined at the beam-level, given as:
    \begin{align}
        C = \frac{\nh(fil) - \nh(bg)}{\nh(bg)}\label{eq:contrast}
    \end{align}
    
   Using the CD map and the skeleton identified, the properties are obtained at the beam level : Filament radius on each side $R_{bg}$, Plummer $p-index$, $R_{flat}$ and linear mass density $\lambda$. The bulk properties obtained at the filament level are : Length, Total mass and the total area.

\section{Result and Discussion}\label{sec:res-dis}

\subsection{Validation: Similarity of filaments via radial profile}\label{sec:val}

     \begin{figure}
        \centering
        \includegraphics[width=0.98\linewidth]{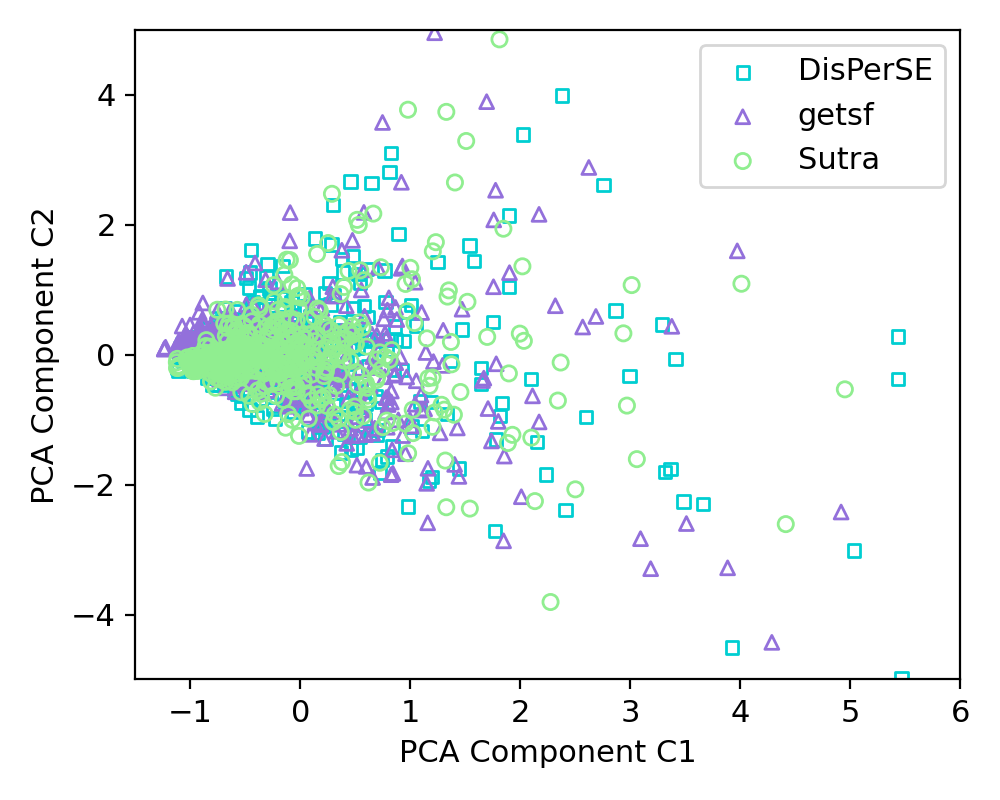}
        \caption{Distribution Components (C1, C2) obtained by PCA of radial profiles of filaments identified by different algorithms \getsf, \disperse, \sutra.}
        \label{fig:pca-comp}
    \end{figure}

    Implementation of \sutra on the fields used in training, (selected chunks containing \ufil skeleton) leads to identification of a large number of new filaments even in the chunks with no \ufil \,filaments (e.g. red areas in Figure \ref{fig:training-skeleton}). Extraction of local filament radial profiles gives scope of additional validation strategy of these newly identified skeleton. If the \unet model is able to learn the underlying pattern that 'defines' the filament, one would expect that new filament would show statistically similar characteristic to the training filaments. To make this comparison, we project the radial profiles onto lower dimension space using principle component analysis (PCA, \cite{pca}). The selection of lower dimension to 2 is based on the PCA reconstruction loss.
    
    Figure \ref{fig:pca-comp} shows the distribution of radial profiles in a two-dimensional space for different algorithms, \disperse, \getsf and \sutra. The distribution appears visually similar for all three algorithms, therefore showing a similarity in the detected structures by all three methods. We further quantify the similarity between the two dimensional distributions by calculating the KL divergence between the distributions:
    
    \begin{table}
        \begin{tabular}{lccr}
            \toprule
            \toprule
                       & \sutra     & \disperse & \getsf \\
            \midrule
            \disperse  & 0.045      &           &        \\
            \getsf     & 0.066      &  0.083    &        \\
            \ufil      & 0.028      &  0.042    & 0.021  \\
            \bottomrule
        \end{tabular}
        \caption{KL divergence between radial profiles extracted using algorithms \disperse, \getsf, \sutra and a \ufil }
        \label{tab:my_label}
    \end{table}

    The low value of KL divergence shows that the filaments extracted by the three algorithms have similar radial profiles. The \ufil skeleton is more similar to \getsf than \disperse. The similarity of \sutra profiles is higher with \ufil skeleton compared to other two methods. This reflects that \sutra has learnt to identify filament crest similar to the \ufil skeleton.

\subsection{Comparative study of \sutra with \getsf and \disperse}
    \begin{figure*}
        \centering
        \includegraphics[width=0.95\linewidth]{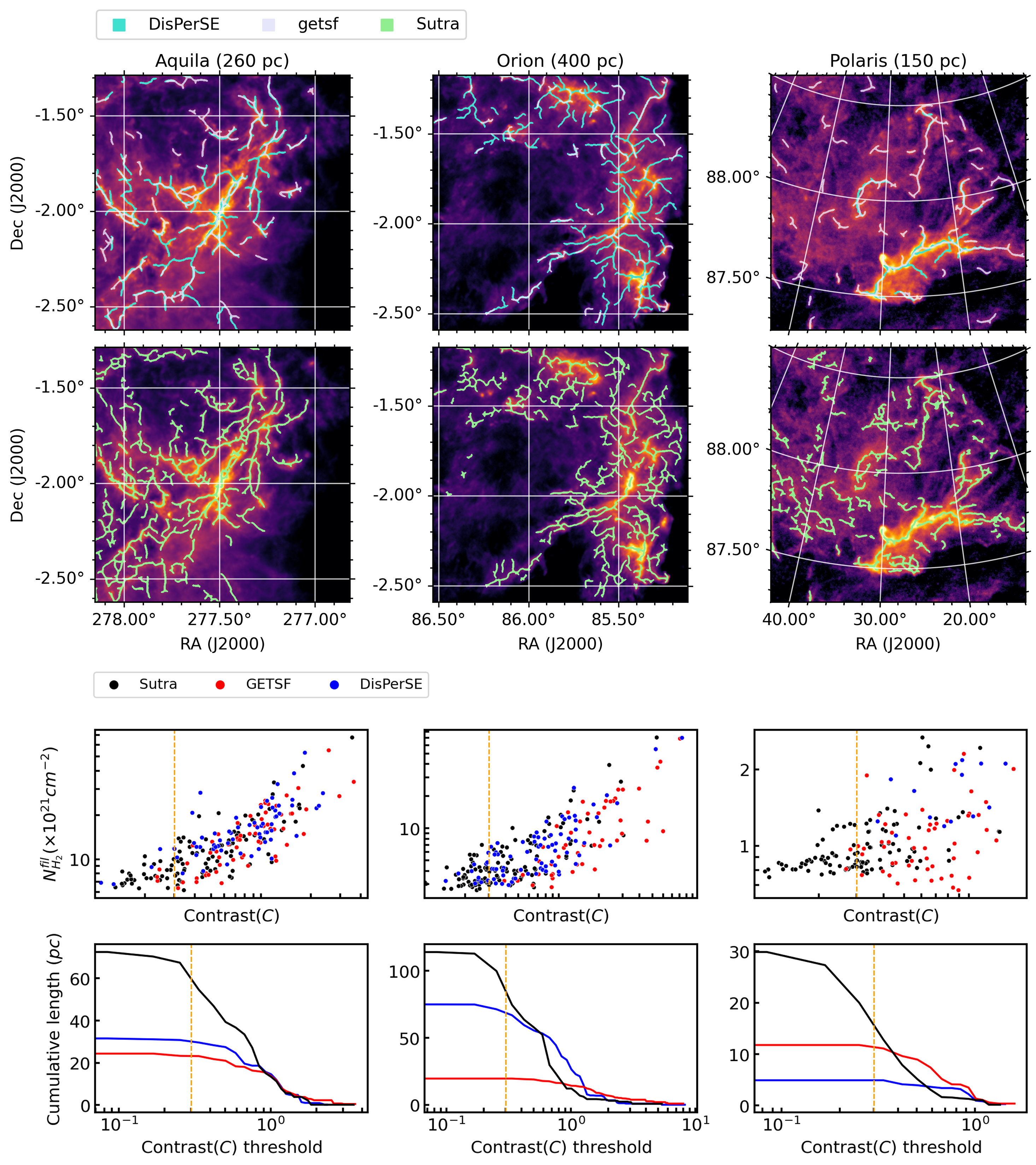}
        \caption{Comparison of \sutra with \disperse and \getsf on cutouts in Aquila, Orion and Polaris clouds (Column-wise). The image plot in top two rows shows comparison of \disperse$+$\getsf filaments(row-1) with the \sutra filaments(row-2). The filaments are thickened for better visualization. In Aquila both \getsf and \disperse filaments mostly overlaps. Orion cutout is mostly dominated by \disperse filaments whereas the Polaris cutout is dominated by \getsf filaments. The filament level variation of filament-mean intensity($\nh^{fil}$) and the total length of filaments detected as a function of the local contrast(eq. \ref{eq:contrast}) as scatter-plot(row-3) and line-plot(row-4) respectively. The length is the total length of filament having contrast above the value given on X-axis.}
        \label{fig:method-comparison-cuts}
    \end{figure*}

    \begin{table*}[ht!]
    \centering
   
    \begin{tabular}{llllllllll}
        \toprule
        \toprule
         Field & Method  & Length & Area & Mass & $\lambda$ & $\nh^{fil}$ & $C$ & $2*R_{flat}$ & $P-index$ \\
         &  & Total & Total & Total & median$\pm \sigma$ & median$\pm \sigma$ & median$\pm \sigma$  & median$\pm \sigma$ & median$\pm \sigma$ \\
         &  & ($pc$) & ($pc^2$) & ($M_\odot$) & ($M_\odot pc^{-1}$) & ($\times 10^{21}cm^{-2}$) &  & ($pc$) &  \\
        (1) & (2) & (3) & (4) & (5) & (6) & (7) & (8) & (9) & (10) \\
        \midrule 
        \multirow[t]{3}{*}{Aquila} & \disperse & 30.82 & 8.25 & 2.58 & 18.39$\pm$10.78 & 15.23$\pm$6.33 & 0.95$\pm$0.47 & 0.11$\pm$0.03 & 2.61$\pm$0.37 \\
             & \getsf & 23.32 & 6.68 & 2.20 & 16.41$\pm$10.64 & 13.94$\pm$7.14 & 0.86$\pm$0.48 & 0.11$\pm$0.02 & 2.68$\pm$0.41 \\
             & \sutra & 60.77 & 15.08 & 3.60 & 10.20$\pm$7.87 & 12.26$\pm$4.56 & 0.60$\pm$0.36 & 0.09$\pm$0.02 & 2.49$\pm$0.29 \\
         \midrule
        
        \multirow[t]{3}{*}{Orion} & \disperse & 70.03 & 27.94 & 6.58 & 9.34$\pm$9.77 & 6.10$\pm$3.18 & 0.77$\pm$0.52 & 0.13$\pm$0.03 & 2.21$\pm$0.40 \\
         & \getsf & 19.55 & 9.44 & 3.44 & 18.18$\pm$18.29 & 7.66$\pm$7.12 & 1.51$\pm$1.26 & 0.14$\pm$0.05 & 2.50$\pm$0.65 \\
         & \sutra & 82.57 & 24.89 & 5.08 & 5.79$\pm$6.07 & 5.47$\pm$3.36 & 0.61$\pm$0.33 & 0.12$\pm$0.03 & 2.20$\pm$0.38 \\
        \midrule
         
        \multirow[t]{3}{*}{Polaris} & \disperse & 4.92 & 0.77 & 0.01 & 1.10$\pm$0.49 & 1.53$\pm$0.59 & 0.85$\pm$0.32 & 0.05$\pm$0.01 & 1.89$\pm$0.26 \\
         & \getsf & 11.58 & 1.81 & 0.02 & 0.57$\pm$0.27 & 1.00$\pm$0.30 & 0.64$\pm$0.31 & 0.05$\pm$0.02 & 1.76$\pm$0.34 \\
         & \sutra & 17.42 & 1.91 & 0.02 & 0.38$\pm$0.16 & 0.98$\pm$0.26 & 0.42$\pm$0.18 & 0.04$\pm$0.01 & 1.76$\pm$0.36 \\
        
        \bottomrule
        \end{tabular}
         \caption{Properties of filaments (with contrast $C>0.3$) identified using \disperse, \getsf and \sutra. Length, Area and Mass (column 3,4 and 5) shows total over all the filaments, `quantifying' the extent of filament. The other columns lists the median properties : linear mass density ($\lambda$), filament intensity $\nh^{fil}(\times 10^{21}cm^{-2})$, contrast ($C$), width ($2\times R_{flat}$) and Plummer fit $p-index$) computed over all the skeleton. The error quoted ($\sigma$) are equivalent standard deviation across the filaments measured using the inter-quantile range ($25\%-75\%$)\citep{iqr}}
    \label{tab:compare-table}
\end{table*}

    To demonstrate the application of \sutra, we use the tool on three cutouts of CD maps of Aquila, Orion and Polaris of $1.33deg\times 1.33deg$ each. The clouds span a wide range of distances (260$pc$, 400$pc$ and 150$pc$ respectively), median CD ($5.9\times 10^{21}cm^{-1}$, $2.1\times 10^{21}cm^{-1}$, $5.8\times 10^{20}cm^{-1}$) and physical conditions. Aquila is a gas-rich star-forming cloud and Orion shows significant star formation activity but Polaris is a quiescent molecular cloud with no star formation activity. For comparative study, we selected regions having sufficient filaments in all three methods and applied identical characterization procedure to each. Figure \ref{fig:method-comparison-cuts} shows the comparative application of \sutra on the three cutouts. Row 1 shows the skeleton identified using \disperse and \getsf, while Row 2 shows the skeleton identified using \sutra. In all three cases, the filaments having length $<3\times HPBW$ and those which fail to fit the Plummer profile are dropped off. Row 1 clearly shows the difference in filament extraction for \disperse and \getsf, especially in low intensity regions. Table \ref{tab:compare-table} lists total summed properties: length, area and mass; and median properties: $\lambda$ (line mass $M_\odot pc^{-1}$), $\nh^{fil}$ (filament CD $\times 10^{21}cm^{-2}$), $C$ (contrast), $2\times R_{flat}$ (filament width, $pc$) and $p-index$ (Power-law index). 

    Across all the three cutouts, \sutra consistently identifies a significantly greater total length and area of filamentary structures compared to both \disperse and \getsf. In Aquila, \sutra identifies the largest total length ($72.5 pc$) of filaments, significantly more than \disperse ($30.82pc$) and \getsf ($23.32pc$). 
    In Orion and Polaris, \disperse and \getsf yield significantly different filament lengths. In Orion, \disperse identified nearly three times more filaments ($70.03pc$) than \getsf ($19.55pc$), while in Polaris \getsf found twice the total filament length ($11.58pc$) compared to \disperse ($4.92pc$). In both cases, \sutra identifies a slightly greater total length of filaments ($82.57pc$ for Orion and $17.42pc$ for Polaris). 
    
    The capability of \unet model in extracting filaments missed by other method is demonstrated by \cite{unetfilZavagno2023} where the U-NET model identified significantly more filaments in the low-contrast low intensity region in the Hi-GAL survey\citep{Molinari_2010}. The U-Net's segmentation and pattern learning capabilities, as well as its training on the union of filaments from the two methods makes \sutra sensitive to a larger population of filaments. This is also seen in a comparison of the median line mass ($\lambda$). Across the three clouds, the median line mass of filaments detected by \sutra is lower than \disperse and \getsf. Along with this, \sutra reports the lowest values for contrast and filament intensity. This shows that \sutra detects fainter and more diffused structures which would be missed by either of the two methods used to generate training sample. \sutra shows a more accurate representation for a low-density cloud like Polaris, where filaments are diffused due to cloud's gravitationally unbound and turbulent nature \citep{turbulantPolaris}.

    Row 3 of Figure \ref{fig:method-comparison-cuts} shows the variation of $\nh$ with the filament contrast. Each scatter point represents an individual filament with a given contrast, $C$ on X-axis and $\nh$ on Y-axis. The vertical line represents the contrast thresholds used by \cite{arz2019}. For Aquila, all three algorithms detect low intensity, low contrast filaments. For Orion, \getsf only detects higher intensity and higher contrast filaments ($C>0.3$). For Polaris, \sutra detects filaments with lower intensity and contrast compared to both \disperse and \getsf. Row 4 shows the total length of identified filaments in the cutout for a given contrast threshold. Each curve shows a flat trend for lower contrast, followed by a declining slope. The declining slope shows that the algorithm is sensitive to filament detection for that contrast value. Between all the three methods, \sutra shows a declining slope starting at a lower contrast threshold, which supports the idea that \sutra is able to detect more diffused and less prominent structures. This indicates that \sutra is not simply finding more of same type of filaments, but is also sampling a broader population of filaments. The lower contrast and lower $\lambda$ filaments represent the sample that are sub-critical with $\lambda < 16\times (T/10K) M_\odot cm^{-1}$ \citep{Hacar_2023} considering median dust temperature $T_{dust} = 15K$ \citep{arz2019}, and are at early evolutionary stages. This capability of \sutra is crucial for understanding the complete lifecycle of filaments, from their initial formation to their role in active star formation.

    \begin{figure}
        \centering
        \includegraphics[width=1\linewidth]{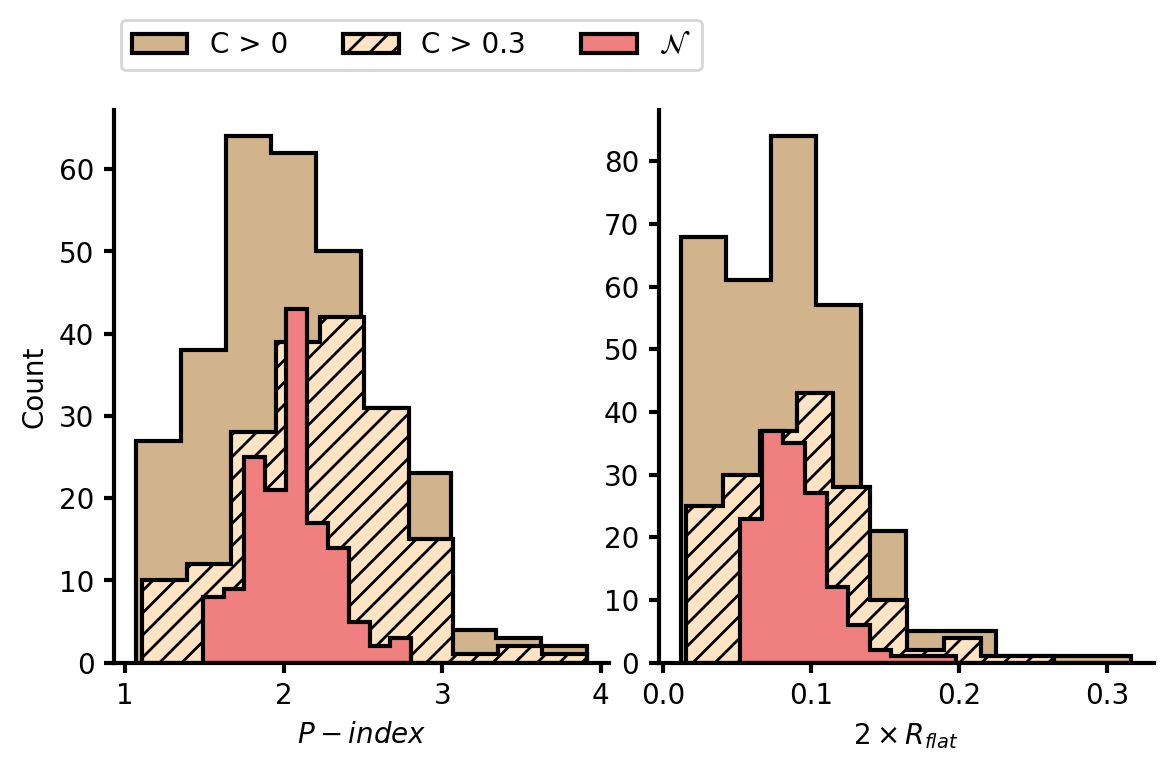}
        \caption{Histogram showing the distribution of $p-index$ and the filament width $2\times R_{flat}$ for all the filaments extracted using \sutra on three cutouts of Aquila, Orion and Polaris (filaments in the bottom panel of Figure \ref{fig:method-comparison-cuts}). \tcm{Histogram corresponding to `$\mathcal{N}$' (\sutra filaments not available in \ufil) is shown in red.}}
        \label{fig:p-r-histogram}
    \end{figure}

    Despite these differences in the overall extent and line density, the median values of intrinsic properties, filament width, ($2\times R_{flat}$) and the $p-index$ shows remarkable consistency across all the three methods (table \ref{tab:compare-table}). The histogram in Figure \ref{fig:p-r-histogram} shows the distribution of $p-index$ and $2\times R_{flat}$ across all the filaments in three cutouts given in Figure \ref{fig:method-comparison-cuts} . The distribution of $p-index$ for $C>0.3$ is similar to the distribution obtained by \cite{arz2019} for 8 molecular clouds at distance $< 500pc$. The typical $2\times R_{flat}$ value around $0.1pc$ agrees with the findings of typical filament width in the nearby molecular clouds \citep{arz2019, andreTypicalWidth}. 
    
    \tcm{To demonstrate that the extra filaments detected by \sutra that not captured in the \ufil skeleton map are meaningful structures, we isolate the portion of the \sutra skeleton that is absent from the \ufil skeleton. Formally, the set of these additional filaments, denoted here as $\mathcal{N}$, is the set-subtraction of \ufil skeleton form \sutra skeleton. Using the physical characterization module (\S\ref{sec:physics-refine}), we perform Plummer fits on $\mathcal N$, obtaining a mean reduced $\chi^2$ of 1.81. Figure~\ref{fig:p-r-histogram} shows the distributions of the fitted $p$-index and the width parameter $2R_{\rm flat}$. We observe that physical properties $p-index$ and $R_{flat}$ follow a similar distribution with  $\langle p-index\rangle = 2.03 \pm 0.25$ and $\langle 2R_{flat}\rangle = 0.09 \pm 0.02$ pc, suggesting that a significant fraction of these structures correspond to physically meaningful filaments.}
    
    The consistency of these intrinsic physical properties suggests that despite the higher detection of low density filaments by \sutra, the physical processes that shape their radial profiles remains the same (see \citealp{Federrath} on how physical conditions affect the width and slope of filament profiles). The consistency also strengthens confidence that \sutra is identifying meaningful structures following cylindrical profiles at a wider range of contrast.

    \tcm{To further assess robustness against background fluctuations and contrast variations, we perform a controlled qualitative comparison using synthetic filamentary maps (Appendix \ref{app:comparison}). In these tests, filaments of varying intrinsic contrast were embedded within fractional Brownian motion (fBm) \citep{fourierISM} turbulent backgrounds of different amplitudes. While Hessian matrix and threshold-driven methods exhibit sensitivity to background amplification and parameter tuning, the \sutra crest-likelihood map remains comparatively stable with respect to absolute filament intensity.}

    \subsubsection{\sutra as a tool to study star-formation}
        \begin{figure}
        \centering
        \includegraphics[width=0.5\textwidth]{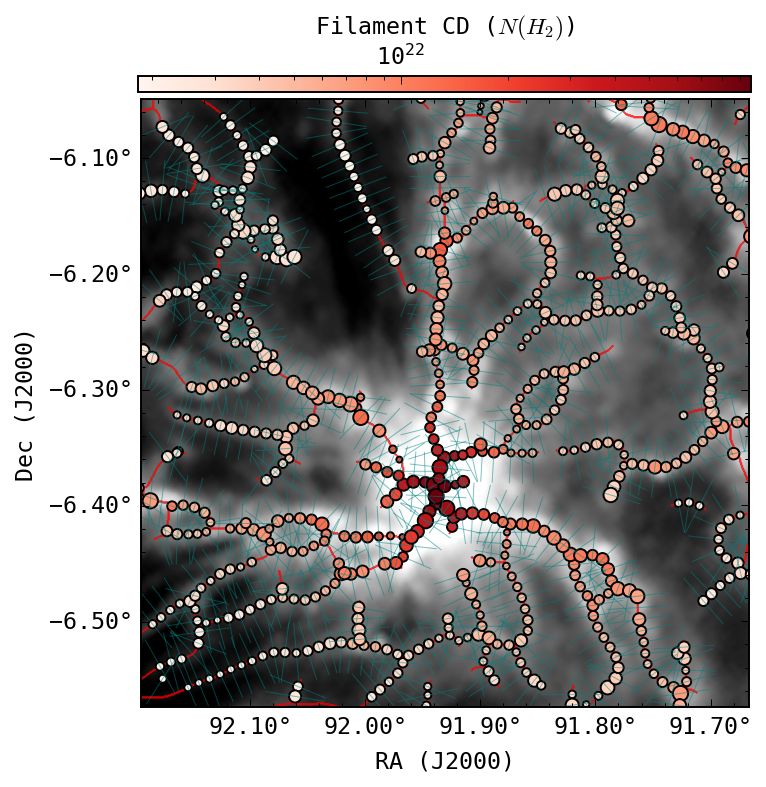}
        \caption{ Example of filament property extraction using \sutra on the Mon-R2 region. Filament skeleton map is shown over the CD map. Variation of filament density variation is shown with 'red' colormap. The radius of beam elements shows the width of filament ($R_{bg}^++R_{bg}^-$)}.
        \label{fig:monr2-example}
    \end{figure}

    \sutra has significant implications for studying the initial conditions of star formation due to its automated filament identification coupled with the ability to perform local, beam-level studies of individual filaments along their axes such as examining how filament properties change as they approach the hub in hub-filament systems \citep{hub2, hub1, hub3, hub4}. To demonstrate this we apply \sutra on Mon R2 CD map \citep{Motte_2010}. The Figure \ref{fig:monr2-example} shows the output of \sutra to generate beam resolved properties map. The variation of filament width and the filament density shows variation as a function of distance from the hub center as seen in the literature \citep{hub4}. This broadens the scope for identifying more intricate filament morphology, including fragmenting filaments exhibiting periodic property variations along their axes \citep{fragmentation} and isolated, `end-dominated` filaments with over-densities at their ends \citep{end1, end2}. This approach yields improved statistical properties compared to traditional methods relying on median profiles \citep{arz2019}.

    Beyond geometric characterization, the derived linear mass densities allow assessment of filament stability by comparison with the critical value expected for isothermal cylindrical filaments. This enables classification of segments as subcritical or supercritical and provides insight into their gravitational state. The availability of beam-level physical parameters therefore allows Sutra outputs to be directly connected to theories of filament evolution and star formation.

    \subsubsection{Limitation and Future Prospects}

    The filament identification module of \sutra, subdivides the input map into chunks. The \unet model then performs segmentation on individual chunks which are joined back together at the end. The \unet model performs parallel segmentation and takes insignificant computational time (wall time $\sim 0.1$s on $1600\times1600$ input CD map of chunk size 64 on Tesla V100 GPU with 32gb VRAM). The current bottleneck in computation time (wall time $\sim 4-5\,minutes$ on $1600\times1600$ input CD map) is the subdivision and then mosaicking of chunks as this step is performed sequentially. This bottleneck time will be crucial for statistical study on a large data sample. This speed can be enhanced exponentially by iterative rejoining of neighboring chunks, where N chunks are converted into N/2 chunks at each iteration. 

    The performance of \sutra's ML model is inherently dependent on the quality and representativeness of its training dataset. Biased dataset can lead to inaccurate or misleading results, particularly when the model is applied outside the scope of training dataset. The current \unet model is trained on \her CD maps of nearby molecular cloud ($d<500pc$) and its applicability on other datasets is out of scope in the present study. Across different surveys, at various length scales and various distances, union of other methods, like template matching \citep{juvela} needs to be included for improving the identification robustness. 

    We emphasize that Sutra presently operates on CD maps only. Deeper physical properties such as filament stability and dynamics require spectroscopic velocity and magnetic field information. Nevertheless, Sutra outputs (line mass, width, crest length) provide the necessary ingredients to combine with molecular line surveys (e.g., CO, NH$_3$) for future studies of gravitational stability, fragmentation, and filament dynamics. We highlight this as a natural next step for applying Sutra in multi-wavelength contexts.
    
    Future prospects include augmentation of \sutra with additional ML models trained on Hi-GAL \citep{higal} and ATLASGAL \citep{atlasgal} data as well as on position-position-velocity data ({\em Kumaran et.al, in prep.}). The modular nature of \sutra, where the user can select the identification model allows for continued refinement and expansion of the training dataset incorporating diverse observational data and simulations for enhancement of the tool's robustness and generalization. \sutra also allows the user to potentially augment the identifier model, representing its significant strength. Furthermore, community contributions for new models, on various datasets, will greatly enhance \sutra's versatility and applicability across the astrophysical community.

\section{Summary and Conclusion}\label{sec:conclusion}

We present \sutra, a unified and modular framework for filament identification and physical characterization. While employing a U-Net backbone for segmentation, the methodological contributions extend beyond conventional mask-based approaches.

The modular design allows straightforward incorporation of alternative segmentation models and facilitates application to diverse datasets. Sutra is released as both a Python package and an online portal to support reproducibility and future extensions.

First, the model is trained directly on single-pixel skeleton maps rather than threshold-derived masks, reframing the task as ridge tracing rather than binary segmentation. Second, training labels are constructed from the union of complementary local (DisPerSE) and global (getsf) filament-finding methods, improving robustness and reducing intensity-dependent bias. Third, beam-level physical characterization is integrated into the pipeline and used as a filtering mechanism to ensure that retained structures exhibit physically meaningful cylindrical profiles.

To do a comparative study we apply \sutra on cutouts of three molecular clouds : Aquila, Orion and Polaris. In all the three cases, \sutra identifies larger extent of filaments following Plummer profile and contrast $C>0.3$. The intrinsic properties $p-index$ and $R_{flat}$ are found to be consistent with the other two methods and in agreement with the known results for the clouds in the HGBS survey. Key implications of \sutra include :
    \begin{itemize}
        \item \sutra traces crest of filament in the CD maps and use the beam-level physical properties to further select robustly identified filaments.
        \item The identification is parameter free and computationally efficient : time taken for identification of skeleton takes time in order of few minutes.
        \item The tool includes integrated characterization algorithm, extracting median radial profiles, which creates filament properties map at beam level, revealing variations in line mass and CD along the filament axes.
        \item The modular nature of \sutra has the potential to augment other ML models and expand its usability to other surveys (Appendix \ref{app:modular}).
    \end{itemize}
    
The \sutra framework produces spatially resolved maps of filament properties along the crest, enabling studies of internal structure and fragmentation. Sutra does not introduce a new network architecture; its strength lies in combining skeleton-based learning, complementary training labels, and beam-scale physical validation within a single, consistent workflow. This integration provides a reproducible pipeline suited for large survey applications.

\tcm{The \sutra framework will be made available as an online portal\footnote{\url{https://github.com/KumaranShivam5/sutra}}, allowing users to upload and analyse targeted CD maps.} In future, the model will have the capability to be tuned for other fields and datasets via transfer learning, making it very adaptive for ISM filament studies. 

\begin{acknowledgments}

We thank the anonymous reviewers for their valuable suggestions and feedback, which have greatly enhanced the quality of this work.

We thank Dr. Nilesh M. Desai (Director SAC) and Dr. Rashmi Sharma (DD-EPSA, SAC) for their support. We extend our gratitude for Tushar Shukla and Pankaj Bodani for their unwavering support and guidance. We acknowledge the feedback from Prof. Anandmayee Tej, Prof. Sarita Vig,  Dr. Manash Samal. 

This research has made use of data from the Herschel Gould Belt survey (HGBS) project \url{http://gouldbelt-herschel.cea.fr}. The HGBS is a Herschel Key Programme jointly carried out by SPIRE Specialist Astronomy Group 3 (SAG 3), scientists of several institutes in the PACS Consortium (CEA Saclay, INAF-IFSI Rome and INAF-Arcetri, KU Leuven, MPIA Heidelberg), and scientists of the Herschel Science Center (HSC).

\end{acknowledgments}

\facilities{\her Space Observatory \citep{Pilbratt_2010}}
\software{\textsc{astropy} \citep{astropy:2022}, \textsc{scipy} \citep{2020SciPy-NMeth}, \textsc{scikit-learn} \citep{scikit-learn}}

\bibliography{compiled_bibliography}{}
\bibliographystyle{aasjournalv7}

\appendix

\section{Comparison of \sutra with other algorithms}\label{app:comparison}

\begin{figure*}
    \centering
    \includegraphics[width=1\linewidth]{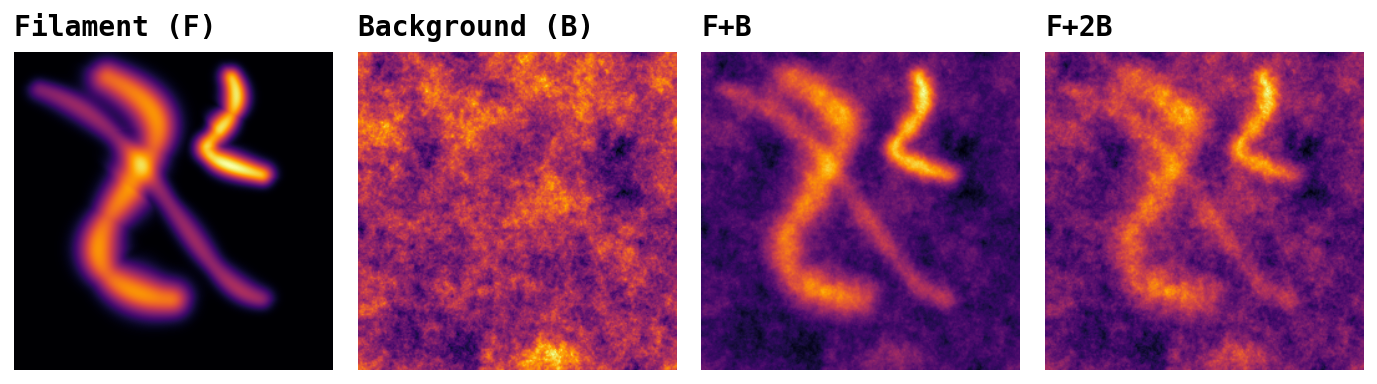}
    \caption{Synthetic data construction used for controlled testing. 
Left: intrinsic filament density distribution ($\mathcal{F}$) generated from analytically defined 3D spines. Second panel: turbulent background density field ($\mathcal{B}$) generated using fBm model with a power-law power spectrum. Third panel: projected column density map of the combined field $\mathcal{F}+\mathcal{B}$. Right panel: higher-background realization $\mathcal{F}+2\times\mathcal{B}$ illustrating reduced contrast.
}

    \label{fig:synthetic-fil}
\end{figure*}
\begin{figure*}
    \centering
    \includegraphics[width=1\linewidth]{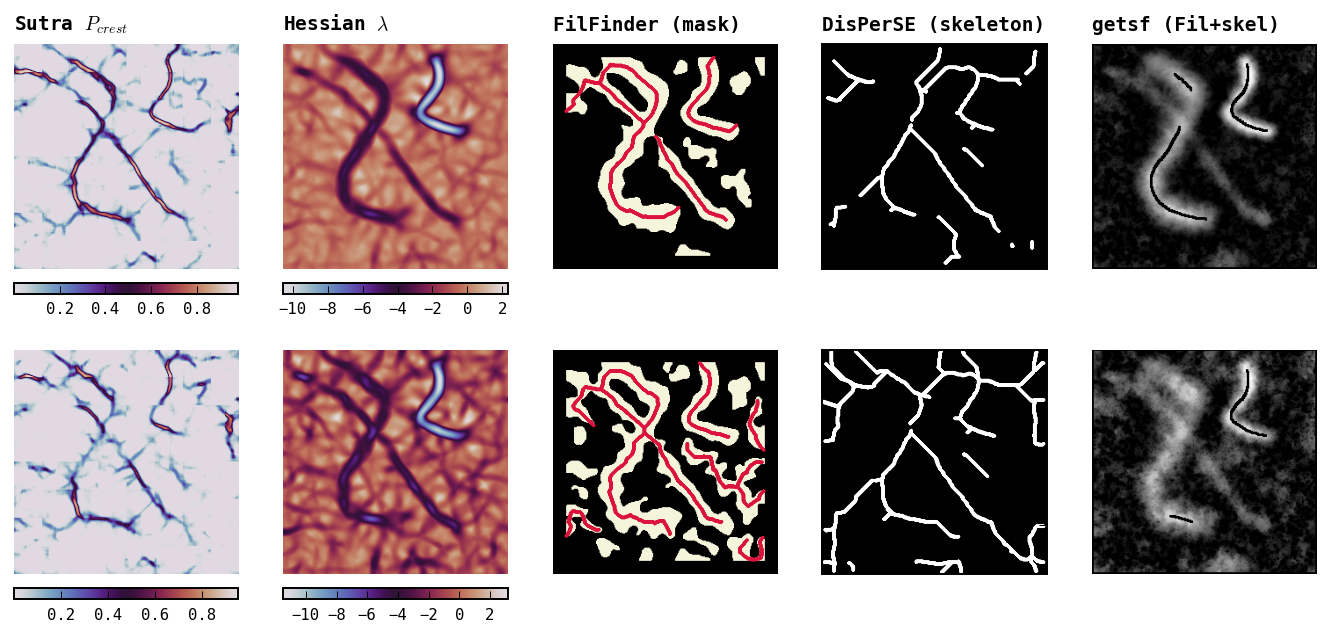}
    \caption{Comparison of structures extracted by different algorithms, tested on a synthetic filament. The top and bottom row shows the result on $\mathcal{F}+\mathcal{B}$ and $\mathcal{F}+2\times\mathcal{B}$ respectively. The panels in Columns shows:( from left to right ):(1) Crest probability map from \sutra, (2) Eigen values of Hessian matrix, (3) mask generated by \filfinder, (4) Skeleton traced by \disperse and (5) Filament and skeleton obtained by \getsf.}
    \label{fig:synthetic-model-op}
\end{figure*}

We perform a controlled qualitative comparison of \sutra with commonly used filament identification methods using synthetic column density maps. This experiment is designed to illustrate relative robustness under varying background levels and dynamic range conditions. It is not used for training, parameter selection, or quantitative benchmarking.

To generate synthetic CD maps, we first define the skeleton of a filament as a 3D Bezier curve, $\mathcal{C}(t)$, where $t \in [0,1]$. We then compute the 3D density distribution, $\rho(x,y,z)$, based on the Euclidean distance $R$ from each voxel to the nearest point on the spine $\mathcal{C}(t)$. we generate three such filaments with central densities as 10,25,50 (arbitrary units). The radial profiles are randomly modulated along axial as well as radial direction to simulate real filaments. 

To simulate a realistic environment we follow fBm model \citep{fourierISM} to generate realistic turbulent molecular cloud environment. We embed the filaments in a log-normal density field with a power-law power spectrum $P(k) \propto k^{-\gamma}$, with $\gamma = 2.1$\citep{gammaMC}. The mean density of background is taken as 5 (arbitrary unit). Both the filament density volume($\mathcal{F}$) and background volume($\mathcal{B}$) are summed to create $\mathcal{F}+\mathcal{B}$ and $\mathcal{F}+2\times\mathcal{B}$ and summed along the Z-axis (Figure \ref{fig:synthetic-fil} panel 3 and 4 respectively). 

Figure~\ref{fig:synthetic-model-op} presents the outputs of different identification methods applied to both realizations. \filfinder parameters were tuned empirically for each case, while \disperse parameters were adopted following \cite{arz2019}. The Hessian-based eigenvalue map ($\lambda$) and the \sutra crest-likelihood map ($P_{\rm crest}$) are shown for comparison.

Both $P_{\rm crest}$ and the Hessian eigenvalue map recover the overall filamentary extent in the moderate background case, though both exhibit difficulty at junction regions. In the amplified background scenario ($\mathcal{F}+2\times\mathcal{B}$), threshold-dependent methods (e.g., \filfinder and \disperse) produce increasing spurious detections. \getsf identifies extended filamentary emission but fails to recover a coherent single-pixel crest in most regions.

\begin{figure}
    \centering
    \includegraphics[width=1\linewidth]{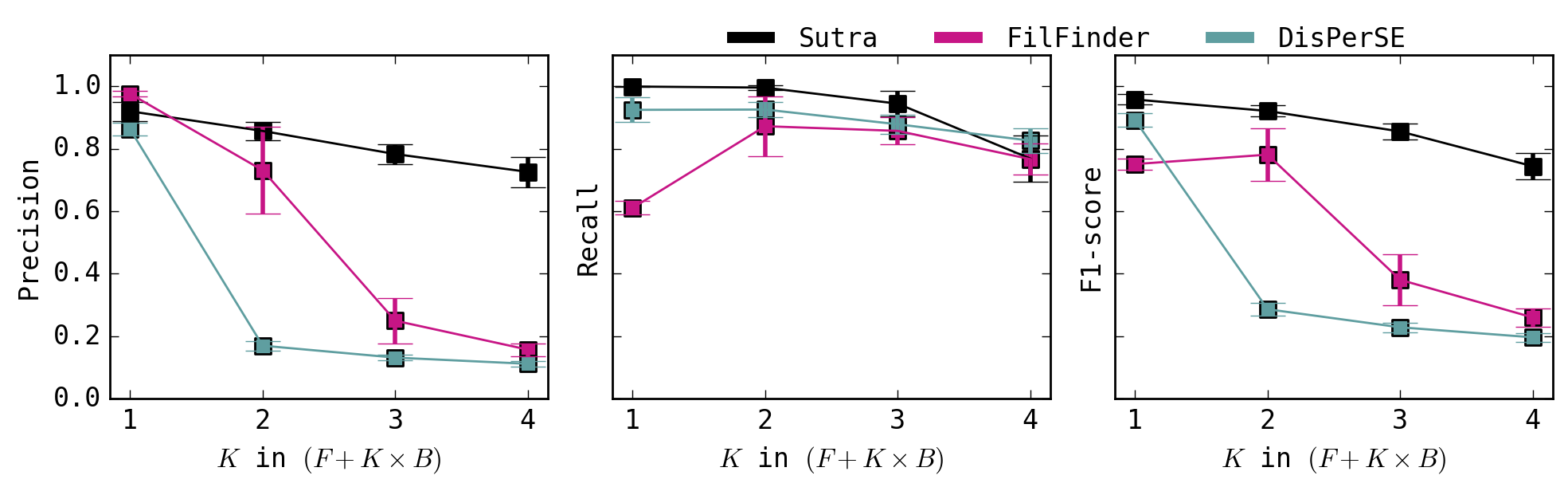}
    \caption{\tcm{Comparison of \sutra, \filfinder and \disperse using precision, recall and F1 score (panels from left to right) on synthetic data with varying background intensity. The error-bar indicates the standard deviation in metric computed on 16 different generated cloud with fixed filament crest for give $K$ on x-axis.}}
    \label{fig:synth-comp}
\end{figure}

A notable distinction is that the \sutra crest-likelihood map remains comparatively insensitive to absolute filament intensity, responding instead to ridge-like geometric structure. This behavior is consistent with the ridge-detection formulation adopted in \S3. \tcm{To complement the qualitative comparison, we perform a quantitative evaluation using the Bezier spine, $\mathcal{C}(t)$ projected on z-axis,  as ground truth reference spine. We consider four background levels, $\mathcal{F}+k\mathcal{B}$ with $k=1,2,3,4$, representing progressively increasing background contamination\footnote{The synthetic column density maps are available on : https://github.com/KumaranShivam5/sutra}. For each k we generate 16 realisations with different random seed. Because a skeleton is a single‑pixel trace of a broader filament, exact pixel‑wise overlap is not expected. A predicted pixel is counted as a true positive (TP) if it lies within 3 pixels of the reference spine (corresponding to an overlap radius of $\tfrac{1}{2}\times$HPBW); otherwise it is a false positive (FP). A pixel in reference spine pixel with no predicted pixel within 3 pixels is counted as a false negative (FN). Precision ($p$), recall ($r$), and the $F1$ score are then computed from these TP, FP, and FN values. $p$, $r$ and $F1$ scores are computed as: }

\begin{align}
    p = \frac{TP}{TP+FP}, \quad r = \frac{TP}{TP+FN}, \quad F1 = \frac{2\times p \times r}{p+r} 
    \label{eq:pr-equation}
\end{align}

\tcm{For comparing \sutra with image threshold based approaches, we perform the same analysis with \filfinder and \disperse. Figure \ref{fig:synth-comp} shows the mean and standard deviation of $p$, $r$, and $F1$ score obtained from 16 cloud realizations spanning a range of background intensities.  No parameter tuning was performed for \sutra with $S_t$ fixed at 0.5 for all cases. For \filfinder the global‑threshold parameter was chosen to maximize the $F1$ score on the baseline cloud $\mathcal{F}+\mathcal{B}$ and then kept unchanged for the other background factors $k$. In the same way, the \textsc{persistence} and \textsc{robustness} parameters of \disperse were optimized on the baseline and left constant for the remaining tests. }

\tcm{We find that \sutra maintains stable performance across increasing background levels, with only a modest decline in precision (from $\sim 0.98$ to $0.8$) and recall (from $\sim 1.0$ to $\sim0.8$). In contrast, both \filfinder and \disperse\ show a significant degradation in precision (from $\sim 0.95$ to $0.2$) at higher background levels due to increased spurious detections, while recall becomes sensitive to parameter choices. In practice, maintaining comparable performance for these methods requires re-adjustment of detection thresholds for each background realization.}

\tcm{This shows a practical advantage that \sutra can be used across varying background conditions without parameter tuning. We emphasize that this comparison is intended to assess robustness rather than to provide a definitive ranking of methods}. The scores obtained on the synthetic clouds serve as a relative performance indicator and should not be interpreted as definitive results for observed CD maps. To translate these metrics to real data, one must first ensure that the synthetic clouds statistically resemble the CD maps of filamentary clouds. A systematic evaluation of various approaches, including non-threshold methods and their parameter sensitivities across diverse cloud conditions for more realistic scenarios is deferred to future work.

\begin{figure}
    \centering
    \includegraphics[width=\textwidth]{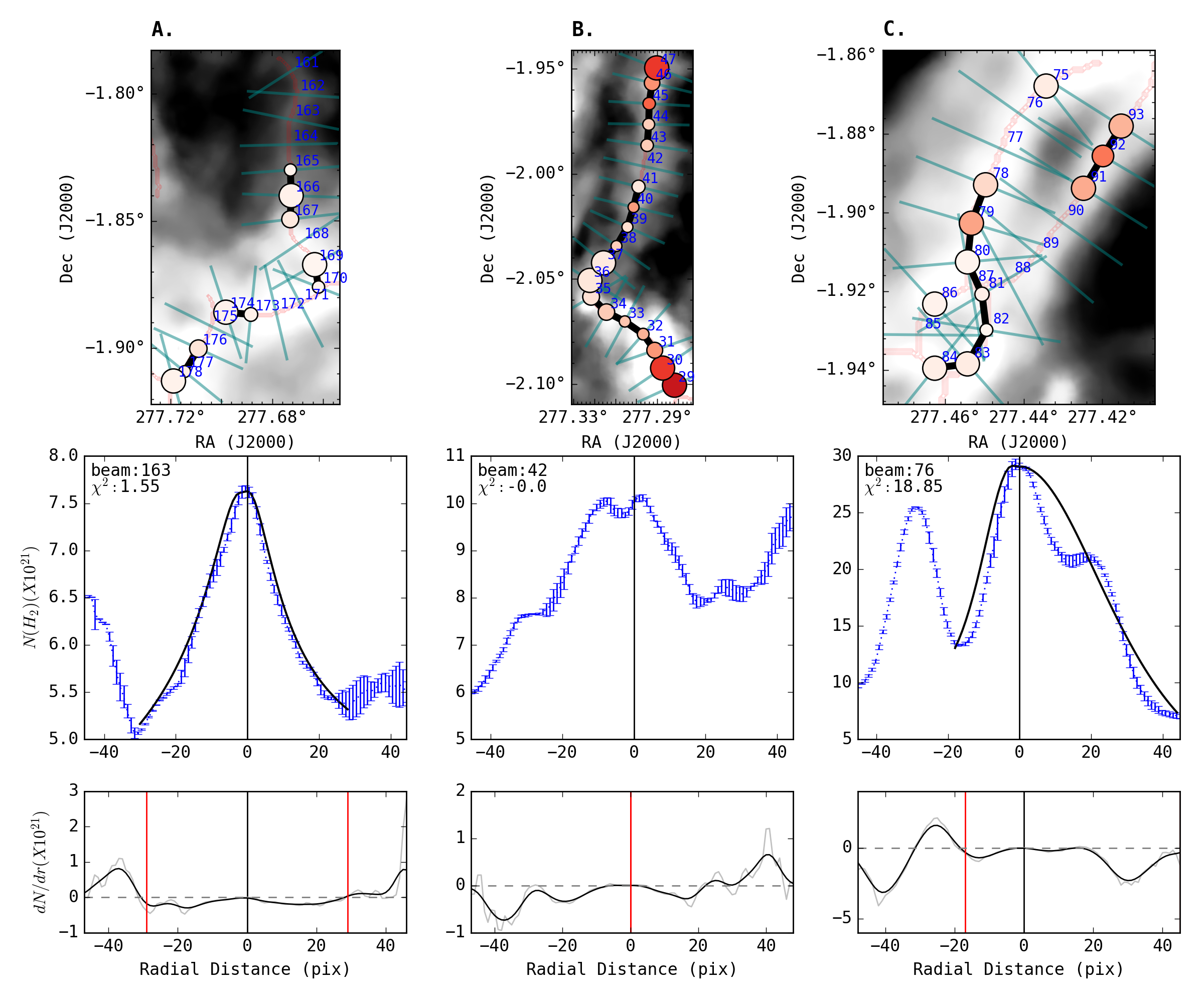}
    \caption{Examples of edge-cases identified by \sutra. The top row shows a single filament from the Aquila molecular cloud. The middle row shows a single beam element from the filament. The bottom row. Panel (A) shows a beam element corresponding to a low-contrast filament. Panel (B) shows a beam element with overlapping filaments, which lead to kinks in the central maxima. Panel (C) shows a filament with overlapping low intensity structures that lead to poor Plummer model fit.}
    \label{fig:sutra-edge}
\end{figure}

\section{Cases for filtering out beam-segments from skeleton map}\label{app:Plummer-filtering}

Along with identification of filaments and extraction of their properties, \sutra is also able to identify the segment of filaments deviating from cylindrical profile. These scenarios represent cases where the radial profile morphology is more complex than the Plummer profile assumed. Examples from three major occurrences are shown in Figure \ref{fig:sutra-edge}. Panel A shows an identified filamentary structure which contains low-contrast regions. The beams at the end of the filament (161-165) have a contrast less than 0.3 with respect to the background. As per the analysis done by \citet{arz2019}, filaments with contrast greater than 0.3 correspond to robust filaments. Therefore, \sutra gives an option to filter out extracted filaments based on contrast.

Panel B shows a beam element, where there is a local minima at the peak of the central intensity. Such cases could correspond to overlapping structures which may be discernible at a higher resolutions. Since this affects the central maxima of the Plummer profile, such beam elements are discarded from further analysis. Panel C shows a complex filamentary region. For the selected beam element, there is an overlapping filament on the left-hand side of the profile, therefore the radial profile is truncated at 18 pixels. However, on the right hand side, there is also a slight increase in CD at 20 pixels. This can be the result of a very low intensity filament overlapping with the brighter filament. Such cases lead to poor fitting, as well as unrealistic values of Plummer parameters, therefore, such beam elements are discarded from further analysis.

\section{Dependence of \sutra output on CD map Pixel Scale}\label{app:pixel-scale}
   \begin{figure}
        \centering
        \includegraphics[width=0.75\textwidth]{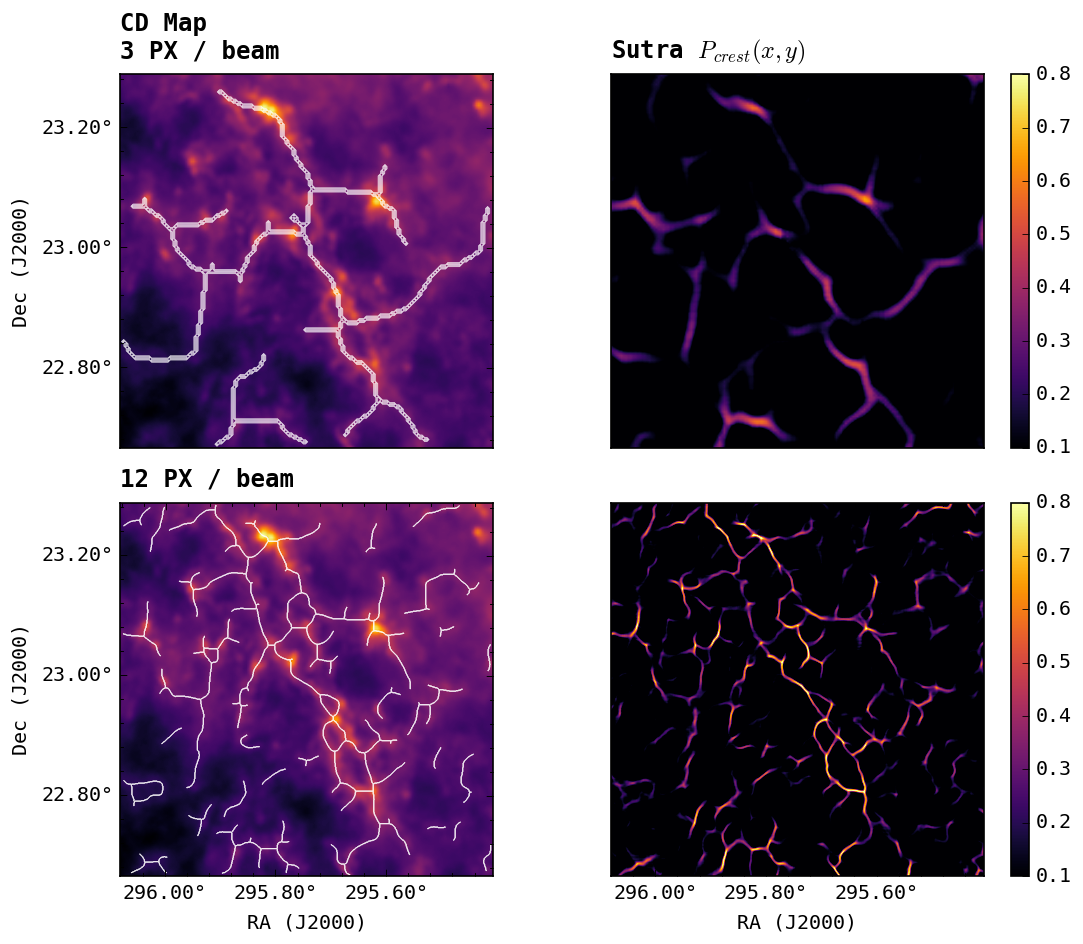}
        \caption{The effect of pixel scale on filament identification by \sutra. The top row shows a CD map at scale 3 pixel/beam and the corresponding probability map identified by \sutra. Note: both the CD maps are convolved to a beam resolution of 36\arcsec before applying \sutra.}
        \label{fig:higal-scale}
    \end{figure}
    Here we demonstrate the impact of pixel scale on Hi-Gal CD map. The CD map is generated at the SPIRE \citep{SPIRE2010A&A...518L...3G} 500$\mu$ resolution. The UN 64 model in \sutra is trained on HGBS survey CD maps \citep{arz2019}. The pixel scale is 3 arcsec/pixel. To apply the trained model on Hi-GAL maps, we regrid the CD map onto a smaller pixel scale (3 arcsec/pixel) keeping the beam resolution same (36\arcsec) using flux-conserving reprojection using \texttt{reproject} package.  Figure \ref{fig:higal-scale} shows the effect of pixel scale on model output crest-probability map. With smaller pixel-scale, the receptive field of UN64 is able to predict finer structure within the beam-resolution limit.

\begin{figure}
        \centering
        \includegraphics[width=1\linewidth]{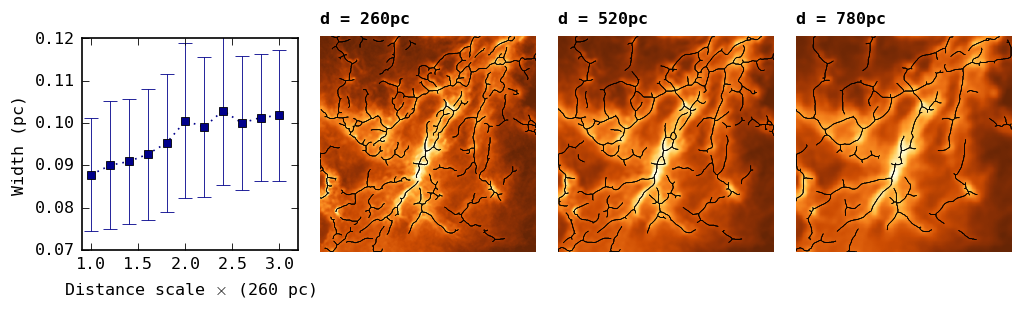}
        \caption{\tcm{Resolution degradation test using effective distance scaling.  \textit{Left:} Median beam-level filament width across as a function of distance (scaled relative to 260\,pc), with error bars indicating dispersion across the sample. \textit{Right:} Example column density maps and corresponding filament skeletons extracted by \sutra at three representative distances (260\,pc, 520\,pc, and 780\,pc). The results show that the derived filament widths remain consistent within expected beam-smearing effects, with a gradual increase at larger distances due to resolution degradation.}}
        \label{fig:dist-scaling}
    \end{figure}

    \tcm{To assess the impact of spatial resolution on filament detection and characterization, we perform an experiment by degrading high-resolution HGBS column density maps. We begin with the $18.2''$ resolution map from \citet{Konyves_2015} and extract filament skeletons using \sutra. Taking the distance of 260pc \citep{Konyves_2015}, we compute the beam-scale filament widths. Further we obtain lower resolution map by convolving it with a Gaussian kernel of $FWHM_s = 18.2\sqrt{s^2 - 1}$, followed by re-gridding to smaller pixel scale, effectively increasing the distance by a factor of $s$. Figure~\ref{fig:dist-scaling} shows the variation of filament width with the effective distance scaling, along with representative examples of the extracted skeletons. The boxplot in the left panel presents the median and errorbars shows the standard deviation of filament widths (computed at the beam-level) as a function of distance. Right three panels show the corresponding maps at selected distances.}
    
    \tcm{The median width at $36.4''$ resolution is  $0.088\pm0.013$ pc, whereas at the lower resolutions $36.4''$ and $54.6''$ the median widths are $0.10\pm0.018$ pc and $0.10\pm0.016 pc$. As expected, higher-resolution maps recover finer filamentary structure, lower-resolution maps smooth out small-scale features. Despite these effects, the filament width remains close to $\sim$0.1\,pc consistent within expected beam-smearing effects. We note that this experiment reflects the combined effect of resolution on both filament identification and subsequent width estimation within the \sutra framework.} Further improvement for application on other surveys can be achieved via the use of transfer-learning, where the first few layers of encoder and last layers of decoder is re-trained on new dataset keeping all the hidden layers' weights and biases frozen.

\section{Modular Architecture of \texttt{Sutra}}
\label{app:modular}

\texttt{Sutra} is implemented both as a Python package and as an online GUI portal. The pipeline is organized into independent, interoperable modules with clearly defined inputs and outputs. Each module can be modified or replaced without affecting the overall framework.

\begin{itemize}

\item \textbf{Filament Identification Module}
\begin{itemize}
    \item Input: column density (CD) map of arbitrary size.
    \item Output: crest-likelihood map ($P_{\rm crest}$).
    \item Preprocessing steps (global and local normalization, contrast transforms, etc.) are implemented as a configurable list of Python functions, each operating as array $\rightarrow$ array.
    \item The trained model is loaded via a dictionary-based configuration of trained weights.
    \item Users can integrate custom trained models without modifying downstream modules.
\end{itemize}

\item \textbf{Skeletonization Module}
\begin{itemize}
    \item Input: CD map and crest-likelihood map.
    \item Operations: probability thresholding, medial-axis refinement, and division into beam-sized segments.
    \item Designed to be model-agnostic; can operate on probability maps from alternative detection algorithms.
\end{itemize}

\item \textbf{Radial Profile Module}
\begin{itemize}
    \item Input: CD map and binary skeleton.
    \item Output: \texttt{RadProf} class object storing radial profiles and fitting results.
    \item Performs beam-scale extraction and Plummer fitting.
    \item Fully independent of the identification module; compatible with externally generated skeletons.
\end{itemize}

\item \textbf{Properties Map Module}
\begin{itemize}
    \item The method \texttt{get\_all\_beam\_props()} evaluates all beam segments and generates \texttt{profGroup} objects containing fitted parameters and profiles.
    \item Maintains a structured dataframe of beam positions (image and Galactic coordinates) and derived physical properties.
    \item Provides hierarchical access to results: entire map, individual filaments, or individual beam elements.
    \item Generates data-table of all beam properties along-with location in Galactic and image pixel coordinates.
\end{itemize}

\end{itemize}

This modular design separates ridge detection, skeleton refinement, radial profiling, and property extraction, enabling extensibility and independent development of each component.

\end{document}